\documentclass[twocolumn, tighten]{aastex62}

\usepackage{apjfonts}
\usepackage{amsmath}

\def\eq#1{\begin{equation} {#1} \end{equation}}
\def\about  {\hbox{$\sim$}}
\def\<#1>{\langle#1\rangle}

\def\N      {\hbox{$\cal N$}}
\def\Ntot   {\hbox{$N_{\rm tot}$}}

\def\twelvCO {$^{12}$CO}
\def\thirtCO {$^{13}$CO}

\def\Dnu    {\hbox{$\Delta\nu$}}
\def\Du     {\hbox{$\Delta u$}}
\def\Dv     {\hbox{$\Delta\v$}}
\def\tauE   {\hbox{$\tau_{\rm E}$}}
\def\tauT   {\hbox{$\tau_{\rm T}$}}
\def\tmin   {\hbox{$\tau_{\rm 0,min}$}}
\def\tmax   {\hbox{$\tau_{\rm 0,max}$}}
\def\numin  {\hbox{$\nu_{\rm min}$}}
\def\numax  {\hbox{$\nu_{\rm max}$}}
\def\Tx     {\hbox{$T_{\rm x}$}}
\def\Tb     {\hbox{$T_{\rm b}$}}
\def\Kline  {\hbox{$K_{\rm line}$}}

\def\Sc    {\hbox{$S_{\rm C\nu}$}}
\def\Ic    {\hbox{$I_{\rm C\nu}$}}

\def\T     {\hbox{$T_\nu$}}

\newcommand{\E}[1]{\hbox{$10^{#1}$}}
\newcommand\lbar  {\hbox{$\bar \ell$}}
\newcommand\lavg  {\hbox{$\<\ell>$}}
\newcommand\Ap    {\hbox{$A_\perp$}}
\let\non=\nonumber

\font\math = cmmi10
\def\m#1{\hbox{\math \char'#1}} 
\def\v{\m{166}}

\font\smallmath = cmmi7
\def\sm#1{\hbox{\smallmath \char'#1}} 
\def\sv{\sm{166}}

\begin{document}

 \shorttitle{Radiation Propagation in Clumpy Media}
 \shortauthors{Conway, Elitzur \& Parra}

\title{\sc Continuum and Spectral Line Radiation\\
  from a Random Clumpy Medium}

\author{John~E.~Conway}
\affiliation{Department of Space, Earth and Environment, 
             Chalmers University of Technology \\
             Onsala Space Observatory, SE-439 92 Onsala, Sweden}

\author{Moshe Elitzur}
\affiliation{Department of Physics and Astronomy, 
             University of Kentucky, Lexington, KY 40506, USA}
\affiliation{Astronomy Department, University of California, 
             Berkeley, CA 94720, USA}

\author{Rodrigo Parra}
\affiliation{European Southern Observatory,  Chile}

 \received{July 15, 2018}
 \revised{August 21, 2018}
 \accepted{August 22, 2018}
 \submitjournal{ApJ}

\begin{abstract}
We present a formalism for continuum and line emission from random clumpy
media together with its application to problems of current interest,
including CO spectral lines from ensembles of clouds and radio emission from
HII regions, supernovae and star-forming regions. For line emission we find
that the effects of clump opacity on observed line ratios can be
indistinguishable from variations of intrinsic line strengths, adding to the
difficulties in determining abundances from line observations. Our formalism
is applicable to arbitrary distributions of cloud properties, provided the
cloud volume filling factor is small; numerical simulations show it to hold
up to filling factors of \about 10\%. We show that irrespective of the
complexity of the cloud ensemble, the radiative effect of clumpiness can be
parametrized at each frequency by a single multiplicative correction to the
overall optical depth; this multiplier is derived from appropriate averaging
over individual cloud properties. Our main finding is that cloud shapes have
only a negligible effect on radiation propagation in clumpy media; the
results of calculations employing point-like clouds are practically
indistinguishable from those for finite-size clouds with arbitrary
geometrical shapes.

\end{abstract}


\section{Introduction}
Radiation propagation in a non-uniform clumpy medium is a common problem in
astrophysics. Example continuum applications include IR dust emission from
circumnuclear tori in active galactic nuclei \citep[AGN;][]{NENKOVA02,
NENKOVA08},  free--free absorption affecting supernova radio light curves and
spectra  \citep{WEILER04}, radio-millimeter wave thermal emission from single
massive stars \citep{IGNACE04} and star-formation induced radio synchrotron
emission accompanied by free-free absorption in galaxies \citep{Lacki13}.
Spectral line applications include modeling the optical and UV spectra from AGN
broad line regions \citep{LAOR06} and interstellar atomic and molecular lines
\citep{MHS84, WALL06, WALL07}. A common approach to the analysis of emission
from clumpy media, pioneered by \cite{MHS84}, is to assume some geometry for
the individual clouds and proceed by averaging over properties along the line
of sight (LOS). Some general scaling relations emerged in these works, but
remained unexplained. In particular, \cite{IGNACE04} noted that, in their
modeling based on spherical clouds, only the distribution in individual cloud
optical depths could affect the spectral shape---the cloud radii were
irrelevant.

In an entirely different approach, \citet{NP84} modeled clumpy media absorption
with point-like identical structureless absorbers, characterized by a single
property, an optical depth, and no other parameters. Noting that random
placement yields a Poisson distribution for the number of absorbers along the
LOS, \citeauthor{NP84} derived the effective optical depth of the medium from
the mean number of absorbers along the LOS and their common optical depth.
\citet{NENKOVA02, NENKOVA08} extended this formalism to the expected emission
from a population of such clouds and placed the \citeauthor{NP84} point-like
absorbers concept on a more solid footing by showing that the ratio of cloud
size to the mean-free-path between clouds is equal to $\phi$, the clouds volume
filling factor. Therefore, when $\phi \ll 1$, each cloud appears as a point
from its nearest neighbor, thus its geometry can be ignored. Still, the usage
of a single optical depth per absorber remained problematic. For example, in
the case of a sphere with optical depth $\tau$ along the diameter, the actual
optical depth along a LOS can vary from $\tau$, for a LOS through the center,
to 0 for a grazing LOS. Here we address this issue, bringing the
\citeauthor{NP84} formalism to completion.

Starting in \S\ref{se:conttrans} we generalize both clumpy absorption and
emission to an arbitrarily complex mixture of clump properties, including
variations of these properties along the LOS. The only restrictions are that
the medium is random (i.e.\ cloud positions are uncorrelated) and that the
propagating radiation does not affect the cloud absorption and emission
properties. We investigate via Monte Carlo simulations the range of volume
filling factors over which our formalism applies and show that significant
departures occur only at relatively large filling factors ($\phi\gtrsim 0.1$;
\S\ref{se:simulations}); introduce the concept of a \emph{clumping factor}
which modulates the  effective opacity of a clumpy medium and depends on the
average properties of its clumps (\S\ref{se:K-factor}); and extend the clump
formalism to include spectral line absorption (\S\ref{se:specabs}) and emission
(\S\ref{se:specemiss}). Section \ref{se:shapes} explores the effects of cloud
shapes and shows that they have no significant impact on radiation propagation
in a clumpy medium. {In \S\ref{se_examples} we apply our formalism to a couple
of current problems involving clumpy emission and absorption by continuum and
spectral lines; Appendix \ref{se:ffabs} provides some additional examples}.
Section \ref{se:discussion} contains a summary and discussion.


\section{Radiation Transfer in Clumpy Media}
\label{se:conttrans}

Consider a region where  matter is concentrated within clouds that occupy a
fraction $\phi$ of the overall volume. The medium will be regarded as clumpy
whenever the filling factor obeys $\phi \ll 1$. This condition is
mathematically equivalent to the requirement that the size of individual clouds
is much smaller than the mean free path between them \citep{NENKOVA08}. In that
case, each cloud can be considered a ``mega-particle''---a point characterized
by its radiative properties but whose shape and size are irrelevant. With this
assumption and taking all clouds to be identical, \citet{NP84} derived the
effective optical depth at frequency $\nu$ along any line of sight through a
clumpy medium in terms of the individual cloud optical depth, $\tau_\nu$, and
the mean number of clouds along the LOS, \N\ (see Appendix
\ref{se_ap:distributions}).

In Appendix \ref{se_ap:contabs} we generalize the Natta \& Panagia result (eq.\
\ref{eq:basic}) to an arbitrary mixture of cloud types and show that radiation
propagating through a clumpy medium will  have a mean transmission
factor\footnote{When averaging over a telescope beam containing many
independent lines of sight sharing the same cloud properties, the measured
value of the transmission factor is expected to lie close to this quantity,
which is formally the statistically average transmission factor for each LOS.}
of $\exp(-\tau_{\rm E \nu})$, where
\begin{equation}
\label{eq:contabs}
         \tauE_\nu = \N\left(1 - \<e^{-\tau_\nu}>\right)
\end{equation}
is the effective optical depth as a function of frequency. In this expression
$\<e^{-\tau_\nu}>$ is the mean transmission factor for radiation passing
through a single cloud, where the averaging is done over all cloud types. This
result reverts to the Nattta \& Panagia expression when all clouds are
identical, but otherwise generalizes it in two fundamental aspects: (1) Even
for point-like, structureless clouds characterized only by optical depth,
$\tau_\nu$ can still vary from cloud to cloud; eq.\ \ref{eq:contabs} shows that
such variations can be handled by simply averaging the transmission factor
$e^{-\tau_\nu}$ over all clouds. (2) Actual clouds cannot really be
characterized by the single parameter $\tau_\nu$. Because real clouds have
geometrical shapes and sizes, the optical depth varies with impact parameter of
the LOS relative to the center of a spherical cloud, the orientation of a
filamentary cloud, etc (see Appendix \ref{se_ap:avg}). Geometrical shape is an
additional cloud property, easily incorporated by adding independent variables
to describe the cloud distribution. Since the average in eq.\ \ref{eq:contabs}
can be made over any number of parameters that describe the cloud population,
when cloud placement is random one can first average over all shape-related
parameters for each cloud type before averaging over the range of cloud types
in the population (for example, with each cloud type perhaps having a different
mean or peak opacity).

The above analysis for absorption is easily generalized to that of emission
from a clumpy medium. \citet{NENKOVA02, NENKOVA08} derive the expression for a
single type of cloud, and Appendix \ref{se_ap:contemiss} generalizes their
analysis to the case of a cloud distribution. The emerging mean emission along
a LOS through a clumpy medium is
\begin{equation}
    \Ic = \int e^{-\tau_{E\nu}(s)} \<\Sc(s)> N(s)ds.
\label{eq:contemiss}
\end{equation}
In this expression $\Sc(s)$ is the source function for single cloud emission at
position $s$ along the LOS and $\<\Sc(s) >$ is its average over all cloud types
at $s$ (see eq.\ \ref{eq:Sav}), $N(s)$ is the mean number of clouds per unit
length at that point (see \S\ref{se_ap:distributions}), and $\tauE_\nu(s)$ is
the effective optical depth, defined as in eq.\ \ref{eq:contabs}, from $s$ to
the edge of the source.

\begin{figure*}
  \centering
 \includegraphics[width=0.75\hsize]{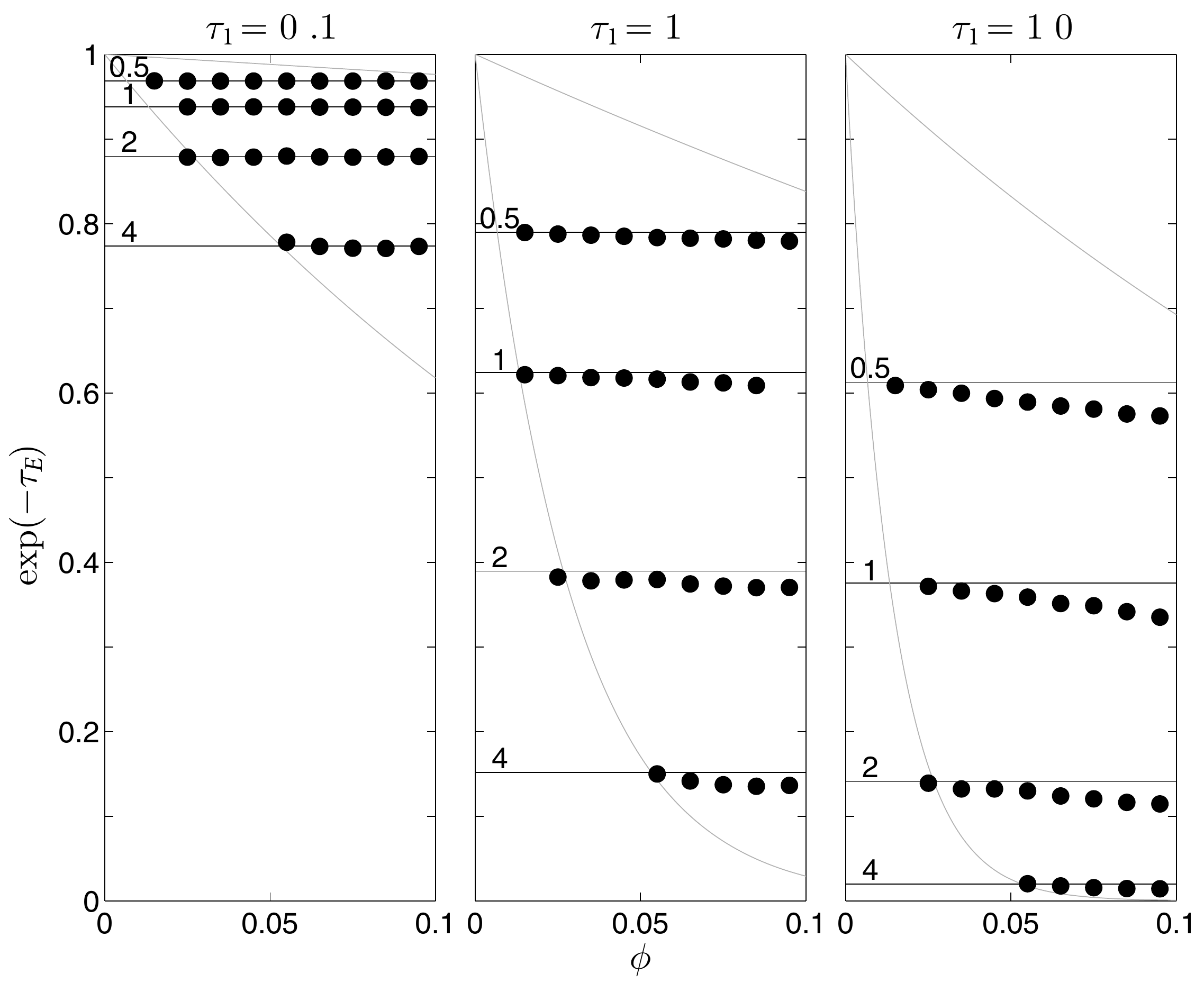}

\caption{Results of Monte-Carlo simulations to investigate departures at large
volume filling factors ($\phi$) from the predictions of the $\phi \to 0$
analytic clump formalism. Each panel shows the expectation value of the
attenuation of background radiation, $\exp(-\tau_{\rm E })$,  by a volume
containing randomly placed identical spherical absorbing clouds. The three
panels show results for clouds with different optical depth $\tau_1$ across the
diameter, as marked. The right panel $\tau_1 = 10$ simulations are
representative of all $\tau_1 \ge 10$. In each panel, horizontal lines show the
$\phi \to 0$  analytic predictions from eq. \ref{eq:contabs} for transmission
factors for differing mean numbers of clouds per LOS, as marked (\N\ = 0.5, 1,
2 and 4). Filled circles show the results of numerical Monte-Carlo simulations
for the transmission factor as a function of $\phi$ for each \N. The statistical
error-bars on these simulations are comparable to the size of the circles. Grey
curved lines trace the boundaries of the simulations; the left one is the locus
of cloud radius 1\% the side-length of the control volume, the right one is
20\% (cf eq.\ \ref{eq:simul}). Significant departures from the $\phi = 0$
analytic expression are seen only for optically thick clouds, and never exceed
$\sim 15\%$ as long as $\phi \le 0.1$.}
 \label{fi:simul}
\end{figure*}

\subsection{Effects of Volume Filling Factor}
\label{se:simulations}

The only properties of the cloud distribution that enter into equations
\ref{eq:contabs} and \ref{eq:contemiss} are the mean number of clouds per LOS,
\N, and the number of clouds per unit length, $N(s)$. There is no dependence on
the cloud volume filling factor $\phi$; by  assuming $\phi \ll 1$ we ended up
with results that are entirely independent of the volume filling factor. The
reason is simple: a complete formalism, in which one would not invoke the
assumption $\phi \ll 1$ from the start, would lead to a series expansion in
powers of $\phi$. The expressions derived above are simply the zeroth order
terms in that expansion, i.e., they are the $\phi \to 0$ limit of a more
complete formalism \citep[see also][]{NENKOVA08}.

Assuming $\phi \ll 1$ at the outset precludes us from studying the effects of
finite volume filling factors and the limitations of our formalism. To try to
quantify the finite-$\phi$ corrections we have carried out numerical Monte
Carlo simulations for the simple case of identical clouds to determine the
critical volume filling factor at which significant deviations from our theory
occur. Our simulations further assume clouds that are spherical and uniform,
and hence fully characterized by a single number, $\tau_1$, the optical depth
across their diameters.\footnote{Since we consider only absorption at a single
frequency, here and in what follows we drop the subscript $\nu$.} We
investigated several values of $\tau_1$, and for each one varied the mean
number of clouds per LOS $\N$ and volume filling factor $\phi$ by filling a
cube with side $2L$ with \Ntot\ spheres of radius $R$ such that
\eq{\label{eq:simul}
    \frac{R}{L} =   \frac{3}{4} \frac{\phi}{\N}, \qquad
    \Ntot=\frac{8}{\pi}\left(\frac{L}{R}\right)^{\!\!2} \N.
}
We carried out a large number of Monte-Carlo simulations, varying $R/L$ and
\Ntot\ (i.e.\ $\N$ and $\phi$) and averaging the results. The simulation
results were then compared with the $\phi \to 0$ theory. Figure~\ref{fi:simul}
presents the comparison, showing that eq.\ \ref{eq:contabs} adequately
describes volume filling factors as large as 10\%, for which the mean
transmission factor $\exp(-\tauE)$ obtained from that expression (see eq.\
\ref{eq:exp_sph}, Appendix \ref{se_ap:avg}) and from the Monte-Carlo
simulations agree to within 15\% at all optical depths. For the particular case
of optically thin clouds, illustrated by the $\tau_1 = 0.1$ case shown in the
left panel, the Monte-Carlo simulations produce close agreement with theory at
all the tested volume filling factors. This is as expected because, as shown
below (\S\ref{se:K-factor}), clumping is irrelevant when individual clouds are
optically thin. In this pseudo-continuous-medium case the effective absorption
depends only on $\N\tau_1$, the overall mean optical depth along the LOS. The
clumpy nature of the medium becomes important only when individual clouds are
optically thick, as seen from the figure's two other panels. In this case
deviations from the analytic results occur at high filling factors, where eq.\
\ref{eq:contabs} over-predicts the fraction of transmitted radiation (i.e.\
under-predicts the effective opacity). The reason is that large volume filling
factors involve large cloud sizes, and keeping the clouds from interpenetrating
each other produces significant deviations from Poisson statistics, leading to
narrower probability distributions for the numbers of clouds per LOS. The
deviations from the analytic $\phi \to 0$ limit are noticeable in the middle
panel, which shows results for $\tau_1 = 1$, and are even more pronounced in
the right panel, where $\tau_1 = 10$. Still, these deviations never exceed 15\%
in any of the simulated cases. Increasing further the single cloud optical
depth has no effect on the outcome because clouds with $\tau_1 = 10$ are
already completely opaque, hence the deviations seen in the figure's right
panel apply to all $\tau_1 \ge 10$.

These results show that eq.\ \ref{eq:contabs} is an excellent approximation for
practical applications. In all cases that we have simulated we find that, up to
$\phi = 0.1$, this analytic expression correctly predicts the absorption by the
cloud ensemble to better that 15\%. We expect that the critical volume filling
factor found in these simulations will also apply to more complex cases
involving a mixture of cloud types. Such mixtures will include values of
$\tau_1$ that are both larger and smaller than 10. As noted above, clouds with
$\tau_1 > 10$ will induce the same deviations from the $\phi = 0$ limit as the
purely $\tau_1 = 10$ case. And since such deviations are even smaller for the
mixture clouds that have $\tau_1 < 10$, the model with $\tau_1 = 10$ for all
clouds provides the maximal deviations from the analytic result in eq.\
\ref{eq:contabs}.

\subsection{The Clumping Correction Factor}
\label{se:K-factor}

Equation \ref{eq:contabs} gives the value of the {\it effective optical depth}
of a clumpy medium $\tau_{\rm E \nu}$, such that $\exp(-\tau_{\rm E \nu})$ is
the expectation or mean value of the transmission of background radiation. In
contrast, the mean of the {\it total optical depth} through the source is
\eq{\label{eq:tauT}
    \tauT_\nu  = \N\,\<\tau_\nu>.
}
This quantity would be the overall optical depth if the clouds were dispersed
into a smooth medium maintaining the same total gas column density along the
LOS. Therefore the ratio
\eq{\label{eq:Kdef}
   K_{\nu} \equiv \frac{\tauE_\nu}{\tauT_\nu}
}
measures the factor by which the clumpy medium effective opacity is reduced
compared to its total opacity due to the  effects of clumping. From equations
\ref{eq:contabs} and \ref{eq:tauT},
 \eq{\label{eq:K}
  K_{\nu} = \frac{1 - \<e^{-\tau_\nu}>}{\<\tau_\nu>}.
}
The functional form of this clumping correction factor resembles the escape
probability familiar from line transfer calculations. When $\tau_{\nu} \ll 1$
for every cloud, i.e., all clouds are optically thin, $K_{\nu} = 1$: the
effective optical depth equals the total optical depth at the given
frequency---clumping is irrelevant. Note that the only requirement for this
condition to be met is that individual clouds be optically thin; the total
optical depth, $\N\,\<\tau_\nu>$, can still be large. If we keep $\tauT_\nu$
fixed and increase the number of clouds \N\ along the path, individual clouds
will decrease in opacity until they become optically thin when $\N >
\tauT_\nu$, yielding $K_{\nu} \simeq 1$; that is, for fixed $\tauT_\nu$ the
absorption properties approach the smooth density limit at large (generally
frequency dependent) \N\ even if the cloud volume filling factor remains small.
On the other hand, when $\tau_{\nu} \gg 1$ for every cloud, $K_{\nu} =
1/\<\tau_\nu>$ and the effective opacity is less than for a smooth medium of
the same mean column density. In this case $\tauE_\nu = \N$, meaning that
although individual clouds are optically thick, a photon still escapes when it
avoids all clouds along the LOS, an event whose probability according to
Poisson statistics is $e^{-\cal N}$.

Since $K_{\nu}$ is always $\le 1$,  clumping can only {\em decrease} the
effective opacity, reflecting the possibility of photon escape even when all
clouds are optically thick. Significantly, $K_{\nu}$ is determined  only by the
mean single cloud optical depth, independent of the number of clouds. Since the
effective opacity obeys $\tauE_\nu = K_{\nu} \tauT_\nu$, we can use $\tauT_\nu$
and $K_{\nu}$ to parameterize a region's absorption properties instead of
$\tau_\nu$ and \N. \emph{Sources with the same $\tauT_\nu$ and $K_{\nu}$ will
have identical absorption properties even if their clouds have very different
shapes and distributions in properties}.

\subsection{Line Absorption by Clouds in Motion}
 \label{se:specabs}

The formalism developed above is applicable to the transfer of radiation at
frequency $\nu$ no matter how the  optical depth $\tau_\nu$ varies with
frequency. Bulk cloud motions introduce Doppler shifts that couple velocity and
frequency distributions. This coupling can be ignored in the case of continuum
radiation, where the variation of $\tau_\nu$ is negligible for any of the
Doppler shifts. Lines are at the opposite limit---a spectral line has $\tau_\nu
= \tau_0\Phi(\nu - \nu_0)$, where $\tau_0$ is the optical depth at the
line-center frequency $\nu_0$ and $\Phi$ is sharply peaked about that
frequency. We now show that thanks to its narrow spectral range, the whole line
can be treated as a single entity rather than frequency by frequency.

Replacing the frequency shift from line center by the equivalent Doppler
velocity \v, the line profile is $\Phi(\v)$, with normalization chosen such
that $\Phi(0) = 1$. The cloud velocity width can then be defined as $\Dv =
\int\Phi(\v)d\v$, a quantity that characterises the range of internal LOS
motions within the cloud.\footnote{A Gaussian  Doppler profile $\Phi(\v) =
\exp(-\v^2/\sigma^2)$  gives  $\Dv = \sqrt{\pi}\sigma$.} Note that the ensemble
of clouds encountered along a LOS can have a range of  different velocity
widths $\Dv$ or line center opacities $\tau_0$.\footnote{Line-center opacities
$\tau_0$ can vary between different kinds of cloud within the ensemble. Even
for identical clouds they can still vary due to encountering a cloud with a
different LOS impact parameter or orientation.} Invariably the observed
ensemble also contains clouds with a range of different bulk Doppler
velocities,  $N(u)$ being the number of clouds encountered on average along the
radiation path per unit LOS velocity at bulk Doppler  velocity $u$ (see
Appendix \ref{se_ap:distributions}). Bulk cloud motions strongly affect the
line propagation when their range is much larger than \Dv. Appendix
\ref{se_ap:specabs} shows that the effective optical depth is given by the
convolution equation
\begin{equation}\label{eq:tauE_line}
       \tauE(\v) = \int N(u)\left(1 - \<e^{-\tau(u - \sv)}>\right) du,
\end{equation}
where the angle brackets denote averages over all cloud properties other than
their LOS velocity.  Since the optical depth across a spectral line profile
vanishes rapidly when $|u - \v| > \Dv$, the contributions to the convolution
integral are confined mostly to a narrow interval with width \about \Dv\ around
the LOS velocity $u = \v$. When $N(u)$ varies only on scales much larger than
\Dv\ it can be taken as constant across the line profile. Then to a good degree
of approximation, the {\it effective optical depth} at \v, the line-center
Doppler shift from rest frequency, is
\begin{align} \label{eq:W}
    \tauE(\v) &= N(\v)\,\<W>(\v),  \\
    \hbox{where}\quad
    \<W>(\v) &= \int\left(1 - \<e^{-\tau(u - \sv)}>\right) du \non
\end{align}
is the velocity equivalent width of the mean normalised emission profile of
single clouds averaged over all cloud types at spectral velocity \v. In this
general case a \v-dependence of $\<W>$ may occur when  typical cloud peak
$\tau_0$ and/or \Dv\ vary with the cloud bulk velocity; for consistency of the
derivation, such variations must occur over velocity scales much larger than
\Dv.  When cloud line shapes and opacities do not depend on the cloud
velocities, the \v-dependence can be dropped and we can consider a single value
of the equivalent width $\<W>$.

Following the definition given in \S\ref{se:K-factor}, the total optical depth
for spectral line emission  at velocity \v\ is, on average, $\int
N(u)\<\tau(u-\v)>du \approx N(\v) \int\<\tau(u-\v>du $. From our definition of
\Dv, $\int\<\tau(u-\v)> du = \<\tau_0\Dv>(\v)$, where the averaging is made
over all properties of the ensemble clouds other than bulk velocity. Therefore
\eq{\label{eq:tauT_line}
    \tauT(\v) = N(\v) \<\tau_0\Dv>(\v).
}
This would be the total optical depth if the absorbing particles in all clouds
along a given LOS were dispersed into a spatially smooth distribution. The
correction factor converting total line opacity into effective clumpy line
opacity is thus
\eq{\label{eq:kspec}
    \Kline(\v) \equiv  \frac{\tauE(\v)}{\tauT(\v)}
                 =     \frac{\<W>(\v) }{\<\tau_0\Dv>(\v)},
}
the line equivalent of the single-frequency clumping factor defined by eq.
\ref{eq:Kdef}. As for the continuum also for line emission: the  effect of an
arbitrarily complex clumpy distribution again is reduced to a multiplicative
factor determined from the properties of individual clouds and independent of
their number. At every spectral velocity \v, the entire line is described by
the single clumping correction factor \Kline(\v). When cloud properties are
independent of bulk velocity, a single multiplicative factor links effective
and total opacities at all line velocities.

\begin{figure*}
  \centering
 \includegraphics[width=\hsize, clip]{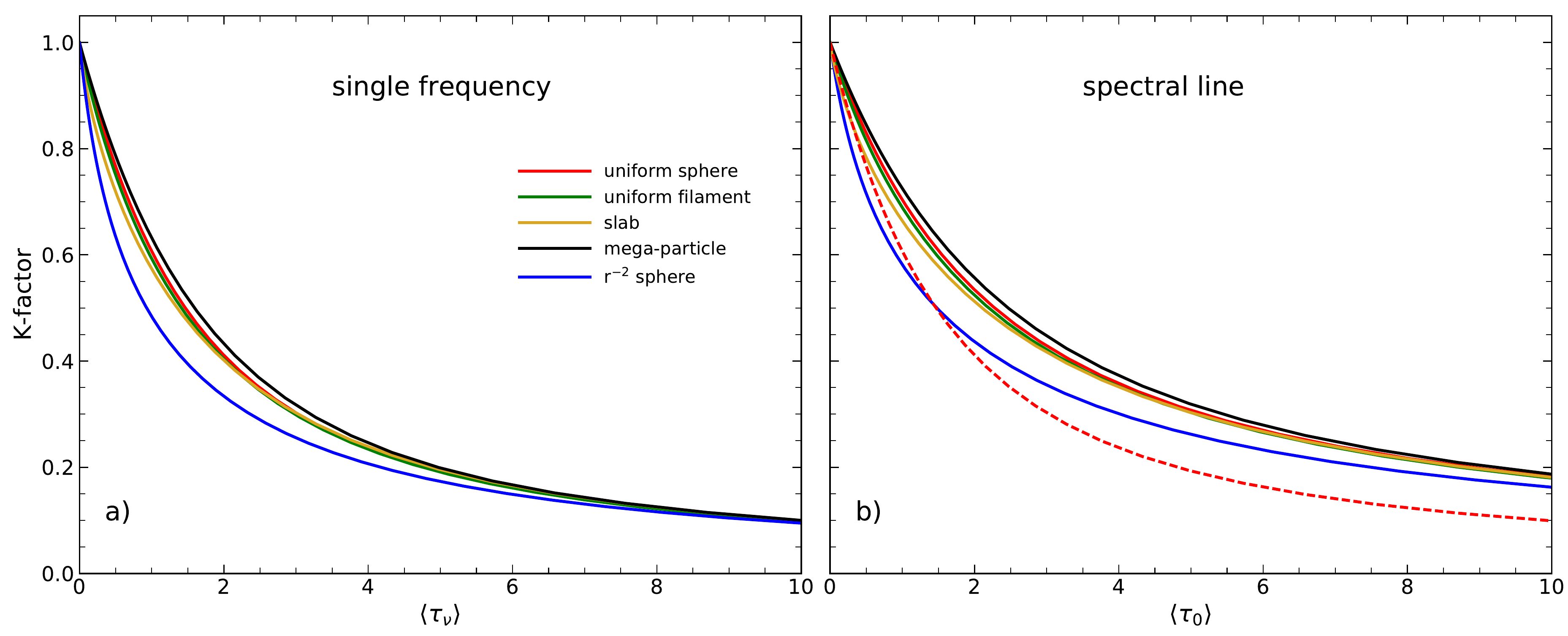}

\caption{Effect of cloud geometrical shape: The clumping correction $K$-factor
as a function of the cloud mean optical depth (\emph{a}) at a single frequency
(eq.\ \ref{eq:K}; relevant to the continuum case) and (\emph{b}) for a spectral
line (eq.\ \ref{eq:kspec}), where $\<\tau_0>$ is the average optical depth at
line center. Different cloud geometries are plotted with different colors, as
labeled; a ``mega-particle'' is a structureless absorber whose only property is
optical depth $\tau_{\nu}$; the $r^{-2}$-sphere is a spherical cloud with
absorption coefficient $\propto r^{-2}$ (see text for details). The dashed line
in the right panel reproduces for comparison the uniform-sphere plot from the
left panel. Note the relatively weak dependence of \hbox{$K$} on geometry,
especially for the slab and the uniform sphere and filament.}

\label{fi:kfactor}
\end{figure*}

When every cloud is optically thin at line center, $\<W>(\v) \approx
\int\<\tau(u-~\v)> du = \<\tau_0\Dv>(\v)$ (eq.\ \ref{eq:W}) so that \Kline\ =
1. When \mbox{$\tau_0 > 1$} for every cloud, the velocity integration that
determines $\<W>$ can be approximated as effectively truncated at $u =
\delta\v_{\tau_0}$, the velocity that separates the optically thick core from
the optically thin line wings. For the Doppler profile $\delta\v_{\tau_0}
\propto \Delta\v\sqrt{\ln \tau_0}$, yielding the large-$\<\tau_0>$ behavior
$\Kline \sim \sqrt{\ln\<\tau_0>}\,/\<\tau_0>$. In contrast, single-frequency
clumping produces $K_\nu \propto 1/\<\tau_\nu>$ at large $\<\tau_\nu>$ (see
\S\ref{se:K-factor}), a steeper decline than that of \Kline\ (see also Fig 2b
below). The reason for the flatter asymptotic behavior of \Kline\ is that
clumpiness has only a negligible effect in the line wings, which are optically
thin even when the cloud is optically thick at the line center. Thus the
overall clumping correction effect is less significant for the line than for
the transfer of continuum radiation at the line-center frequency through a
clumpy medium with the same total optical depth.

\subsection{Spectral Line Emission}\label{se:specemiss}

The emission from a population of line emitting clouds can be calculated
following the steps outlined for the continuum case in  \S\ref{se:conttrans}.
The details are given in appendix \ref{se_ap:spec_em}, where it is shown that
eq.\ \ref{eq:contemiss} is applicable as is, in unmodified form,  to the case
of line intensity emerging from a clumpy medium. This relation is the
equivalent of the standard formal solution of the radiative transfer equation
for clumpy media, with the appropriate attenuation term for continuum and line
radiation. The only slight complication for the lines is the two-step averaging
in eq.\ \ref{eq:line_S}, involving also averaging over the spectral shape of
single cloud emission (eq.\ \ref{eq:Psi}) to get the local average cloud source
function. Equation \ref{eq:contemiss} holds at every frequency/velocity  for
line radiation whenever the ensemble averages for the cloud emission and
absorption properties are independent of each other, i.e., when the optical
depths of individual clouds are unaffected by the propagating line radiation.
When this condition is violated, the intensity calculation requires an
iterative procedure based on the expressions given here. Details of an
iteration scheme analogous to standard $\Lambda$-iterations are described in
\cite{NENKOVA08} (see \S3.2 of that paper).

When we express intensity in brightness temperature units and use the common
approximation of a constant line excitation temperature \Tx\ inside each cloud,
the brightness temperature \Tb\ of the emergent radiation assumes the simple
form of eq.\ \ref{eq:Tb}. When the cloud velocity distribution is much broader
than the velocity width of internal cloud motions this becomes
\begin{eqnarray}
\label{eq:simpspec}
  \Tb(\v) &=& \<T_{x}> \left[1 - \exp(-\<W> N(\v))\right]\\ \non
          &=& \<T_{x}> \left[1 - \exp(-\Kline \tauT(\v))\right],
\end{eqnarray}
given in terms of either the mean cloud equivalent width $\<W>$ and the number
of clouds per unit velocity, $N(\v)$, or the clumping correction factor \Kline\
and the total optical depth versus velocity, $\tauT(\v)$.

\section{Effects of Cloud Geometry}
\label{se:shapes}

Among the many variables affecting the radiative properties of a clumpy medium
is the geometrical shape of the clouds. Our formalism shows that the shape
affects radiation attenuation only through the clumping factor $K$, enabling us
to study the impact of geometry. To separate the effect of shape from all other
properties we consider an ensemble of identical clouds. When the clouds are
additionally taken to be uniform, the required averages involve only variation
of length along the LOS through the cloud, and thus depend on geometry alone
and nothing else. Since a sphere is the ultimate isotropic shape while an
elongated filament and a semi-infinite slab are at the other extreme end of
anisotropy, these geometries can be expected to bracket the full range of
$K$-factor variation that cloud shapes can induce. Note, in particular, that in
studies of AGN, the slab geometry is invoked in all calculations of broad-line
emission \citep[see, e.g.,][]{LAOR06} and in some models of torus IR emission
\citep{NENKOVA08}.

The averages involve integrations over the cloud projected area for a sphere,
orientation for a slab and both of them in the case of a filament; Appendix
\ref{se_ap:avg} gives the details. The left panel of figure \ref{fi:kfactor}
shows the variation of $K_\nu$ with $\<\tau_\nu>$, the single-frequency cloud
mean optical depth relevant for continuum emission, for slabs (orange curves)
and for uniform spheres (red) and filaments (green). The three are hardly
distinguishable from each other. This weak dependence on cloud geometrical
shape is easy to understand: As noted above (\S\ref{se:K-factor}), irrespective
of geometry $K$ is \about 1 at small optical depths and $\sim 1/\<\tau>$ at
large ones. The geometrical shape affects only the transition between these two
regimes, where its impact is constrained by the limits that must be joined on
either end; Appendix \ref{se_ap:avg} provides some further insight into this
behavior.

The similarity of the results for slabs, spheres and filaments suggests that
all uniform clouds, whatever their geometry, have roughly the same $K$-factor
when taken as a function of $\<\tau>$. To further test this possibility we
considered the extreme limit of ``mega-particles''---structureless clouds whose
only property is optical depth $\tau_\nu$; then $\<\tau_\nu> = \tau_\nu$ and
$K_\nu = (1 - e^{-\tau_\nu})/\tau_\nu$. Shown in black in figure
\ref{fi:kfactor}, the $K$-factor for a mega-particle closely resembles that for
a sphere; the largest difference between the two is less than 5\% (at $\<\tau>$
= 2.2).

These results show that the cloud shape plays only a minor role for uniform
clouds. Internal structure within clouds adds another degree of freedom. To
gauge its potential impact, the blue curve in Figure  \ref{fi:kfactor} shows
$K_\nu$ for spherical clouds with absorption coefficient proportional to
$r^{-2}$ between radii $r_{\rm min} \le r \le 10r_{\rm min}$. While this curve
is significantly outside the narrow range covered by uniform clouds, still the
largest difference between the $r^{-2}$-sphere and a uniform one is no more
than \about 20\% (at $\<\tau>$ = 1.1).

The right panel of figure \ref{fi:kfactor} shows the spectral line clumping
factor \Kline\ (eq.\ \ref{eq:kspec}). For comparison, the single-frequency
$K_\nu$ for a uniform sphere is reproduced with a dashed line. At the same
$\<\tau>$, \Kline\ is significantly larger than $K_\nu$. The reason is that,
whatever the optical depth at line center, the wings of a spectral line always
become optically thin at some point and the effect of clumping disappears
there. In addition, \Kline\ exhibits a flatter decline than $K_\nu$ at high
optical depths---as expected, given that the large-$\<\tau>$ asymptotic
behavior of $\Kline$ is $\sqrt{\ln\<\tau_0>}\,/\<\tau_0>$, while that of
$K_\nu$ is $1/\<\tau_\nu>$ (see \S\ref{se:specabs}). Apart from these
differences, the overall behaviour of \Kline\ closely resembles that of
$K_\nu$, similarly displaying only a weak dependence on geometry. \emph{Cloud
shape is of secondary importance to radiation propagation in clumpy media}.


\section{Example applications}
\label{se_examples}

{In this section we illustrate our general formalism for clumpy media by
applying it to two concrete cases, involving continuum and line emission. Some
additional examples are provided in Appendix \ref{se:ffabs}}

\subsection{Radio Continuum Emission from Clumpy  \\
             Ultra-Compact HII Regions}
\label{se:UCHII}

Radio free-free emission from ultra-compact HII regions (UCHII) often has a
power-law spectrum, $I_\nu \propto \nu^\alpha$, with spectral index $\alpha
\sim 1$ over a wide range of frequencies \citep[][hereafter IC04]{IGNACE04}.
Such spectra are too flat to be explained by optically thick free-free emission
($\alpha=2$), too steep to be optically thin ($\alpha=-0.1$) and extend over
too wide a frequency range to be the transition between optically thin and
thick emission from a single gas phase. Although smooth ionized outflow models
can explain the observed spectra, IC04 argue that the wind properties become
physically implausible for $\alpha \geq 1$. Therefore IC04 instead modeled such
radio--millimeter spectra by a clumpy medium in which the clumps had a wide
distribution of emission measures. Specifically, they investigated the emission
from an assembly of uniform spherical clouds. The free--free optical depth
across the diameter of each cloud is
\eq{\label{eq:tauff}
    \tau_\nu = \tau_0\,\left(\frac{\nu_0}{\nu}\right)^{2.1}
}
where $\tau_0$ is the optical depth at some fiducial frequency $\nu_{0}$. The
cloud number distribution as a function of $\tau_0$ was taken as a truncated
power law between two limits:
\eq{\label{eq:Nff}
    N(\tau_{0}) \propto \tau_0^{-\gamma}, \qquad
    \tau_{0,\rm min} \le \tau_0 \le \tau_{0,\rm max}.
}
As expected, IC04 found that the cloud ensemble spectrum had $\alpha = 2$ at
very low frequencies, where all clouds were optically thick, while at very high
frequencies, where all clouds were optically thin, the spectral index was
$\alpha = -0.1$. In between, over a broad region of frequencies, in which some
clouds were optically thick and some thin, there was an almost constant
intermediate spectral index, $\alpha \approx 1$. \cite{IGNACE04} found they
could fit the spectra observed in UCHII regions assuming $\gamma \approx 1.5$
and a large ratio $\tau_{0,\rm max}/\tau_{0, \rm min} > 100$. However, their
computed spectra assumed that the overall mean number of clouds per LOS obeyed
$\N \ll 1$; that is, they did not treat the effects of one cloud shadowing
another.

\begin{figure}
\includegraphics[width=\hsize]{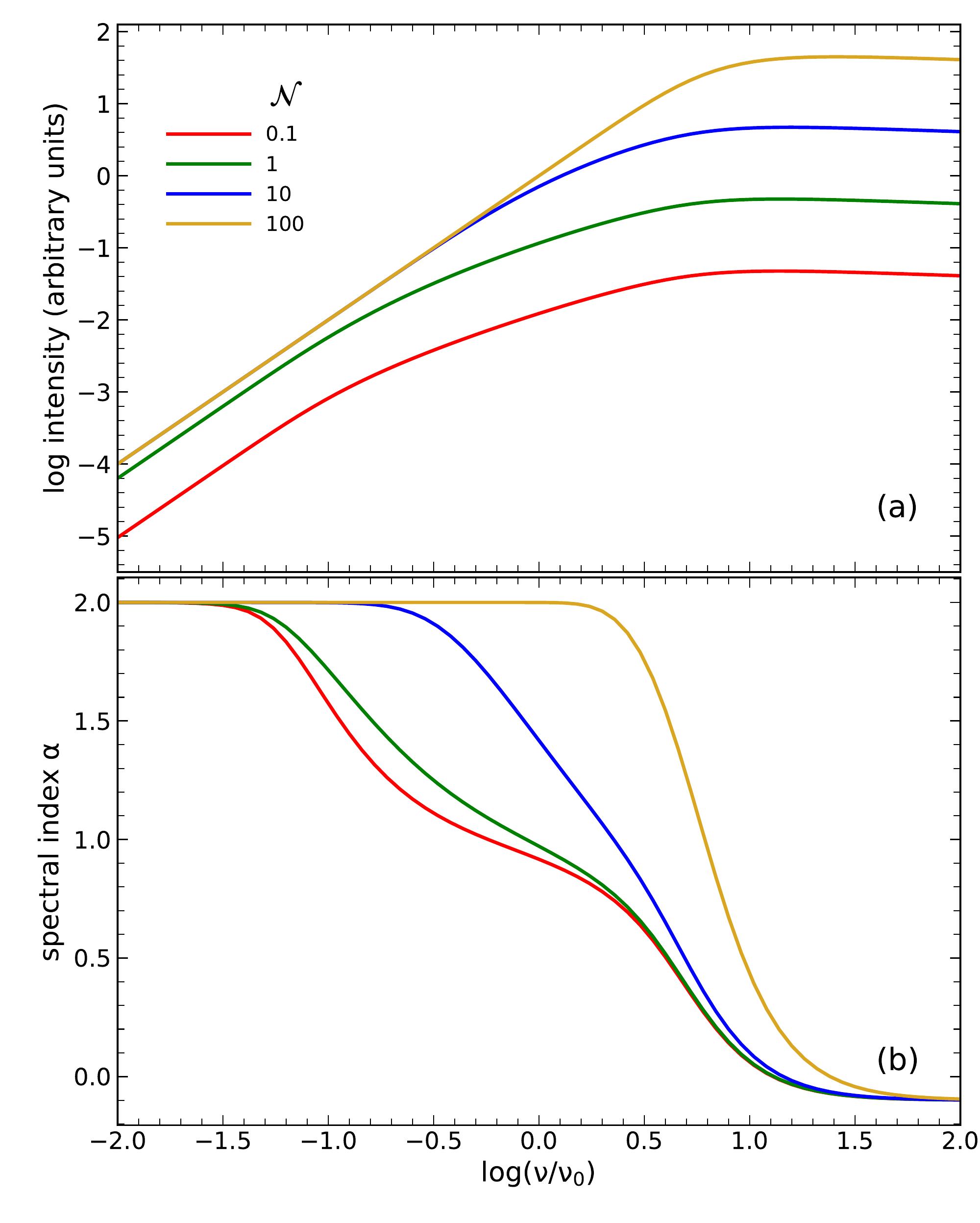}

\caption{Free-free emission from a cloud ensemble with a wide range of optical
depths in a truncated power law distribution (eq.\ \ref{eq:Nff}). ({\em a})
Emerging spectra for various values of \N, the mean number of clouds along the
LOS, as marked, and ({\em b}) the corresponding spectral index $\alpha = d\ln
I_\nu/d\ln\nu$. The resulting spectral indices span the range between the
optically thick ($\alpha = 2$) and thin ($\alpha = -0.1$) limits for single
cloud emission. The \mbox{\N\ = 0.1} case (red curves), which is representative
of all $\N \ll 1$, shows a significant frequency range over which $\alpha$ is
close to unity, but this range narrows as \N\ increases.} \label{fi:hiispec}
\end{figure}

While IC04 argued that cloud shadowing was of minor importance for the one
source they modelled in detail (W49N B2, where $\N \approx 0.15$), it is
clearly of interest to generalize their results to all values of \N. This can
be readily done using the formalism developed here. The general expression for
continuum emission from a clumpy medium is given by equation
\ref{eq:contemiss}. When both the mean cloud emissivity $\<\Sc>$ and
transmission factor $\<e^{-\tau_\nu}>$ are independent of position along the
LOS, it is convenient to introduce the variable $t = \N(s)/\N$, the fraction of
the total number of clouds along the LOS to position $s$ in the source (see
Appendix \ref{se_ap:distributions}). Then $N(s)ds = \N dt$ and the
$t$-integration is immediate, yielding
\begin{equation} \label{eq:hiispec}
  I_{\rm C \nu} = \<\Sc>\,
        \frac{1 - e^{-\tau_{\rm E\nu}}}{1 - \<e^{-\tau_\nu}>}
\end{equation}
where $\tauE_\nu$ is given by equation \ref{eq:contabs}. With this result we
can calculate the emerging spectrum from an ensemble of uniform spherical
clouds similar to that investigated by IC04, but now removing the limitation
$\N < 1$.

We assume that each cloud emissivity is given by the Planck function
$B_\nu(T)$, where $T$ is the cloud temperature, then the cloud contribution to
the specific intensity of the ensemble is $B_\nu(T)\left(1 -
e^{-\tau_\nu}\right)$, where $\tau_\nu$ is the cloud optical depth along the
impact parameter corresponding to a particular LOS. To form $\<\Sc>$ we must
average over the cloud ensemble, i.e., cloud opacities and temperatures as well
as LOS impact parameter. Assuming that $T$ and $\tau_0$ are uncorrelated and
utilizing the Rayleigh-Jeans limit for $B_\nu$ in the radio regime yields
$\<\Sc> = \left[2k\<T>\nu^2/c^2\right] \left( 1 - \<e^{-\tau_\nu}>\right)$. In
this expression the average for the self-absorption factor is taken over both
the cloud LOS impact parameter and the $\tau_0$ distribution. Substituting in
eq.\ \ref{eq:hiispec} we get
\begin{equation}
  I_{\rm C \nu} = \frac{2k\<T> \nu^2}{c^2} \big(1 - e^{-\tau_{\rm E\nu}}\big).
\label{eq:hiispec2}
\end{equation}
This is the same expression as for emission from a single cloud, except that
the cloud temperature is replaced by the ensemble average $\<T>$ and its
optical depth $\tau_\nu$ is replaced by $\tauE_\nu$, the effective optical
depth of the cloud distribution at each frequency. This latter quantity in turn
is given by equation \ref{eq:contabs}, which can be written $\tauE_\nu = \N
f(\nu)$,  where \hbox{$f(\nu) \equiv 1 - \<e^{-\tau_\nu}>$} is the average
absorption fraction of radiation at frequency $\nu$ encountering a cloud in the
distribution; this quantity is determined purely by the mean properties of
individual clouds and is independent of the number of clouds.

Using eq.\ \ref{eq:hiispec2} we can estimate for arbitrary $\N$ the emerging
spectrum from a clump population  containing a wide range of optical depths.
Taking uniform spheres for the individual clouds, the required averaging
carried out in forming the frequency profile $f(\nu) = 1 - \<e^{-\tau_\nu}>$
can first be made over different clump LOS impact parameters (see Appendix
\ref{se_ap:avg}) and then these spatial averages themselves are averaged over
the population of clump types with different opacities across their
diameters.\footnote{In the example chosen below with a wide cloud opacity
distribution, the average over clump structure has negligible effect on the
final spectrum  compared to the effects of averaging over the opacity
spectrum.}

In Figure~\ref{fi:hiispec}, the top panel (a) shows the emerging spectra for
different mean numbers $\N$ of clumps per LOS, the bottom panel (b) shows the
corresponding  local spectral index  $\alpha = d\log I_\nu/d\log \nu$ versus
frequency for each \N. The spectra were calculated with the same power law
distribution of clump opacities assumed by IC04, i.e, the $\tau_0$-distribution
given by  eq.\ \ref{eq:Nff} with $\gamma = 1.5$, \tmin\ = 0.01 and \tmax\ =
100.\footnote{The effects of varying all parameters other than \N\ have been
studied extensively in IC04.}  Given that $f(\nu) < 1$ (from its definition),
it follows that when $\N \ll 1$, all frequencies obey $\tauE_\nu = \N f(\nu)
\ll~1$, in turn implying that $ I_{\rm C \nu} \approx 2 \N k\<T> \nu^2
f(\nu)/c^2$; that is, the output spectrum has a fixed shape $\nu^2 f(\nu)$
scaled in intensity by $\N$. This low-$\N$ limit is illustrated by the red
curve in Figure~\ref{fi:hiispec}(a) for the specific case $\N= 0.1$; this is
the output spectrum found by IC04 and it includes a broad frequency range with
intermediate spectral index. This range starts at the frequency at which every
cloud in the opacity distribution is optically thick, $\numin = \nu_0
{\tau_{\rm 0,min}}^{1/2.1}$, and ends at the frequency at which all are
optically thin, $\numax = \nu_0 \tau_{\rm 0,max}^{1/2.1}$. The spectral index
$\alpha$ of the $\N < 1$ output emission spectrum is therefore intermediate
between its optically thick and thin values over the fractional frequency range
\eq{\label{eq:range}
    \frac{\numax}{\numin} = \left(\frac{\tmax}{\tmin}\right)^{\!\!0.48},
}
which is about 100 for the ratio \tmax/\tmin\ = \E4\ employed here. It is
interesting to note that this relative spectral width is independent of the
absolute values of the optical depth boundaries; changing the lower limit of
the $\tau_0$-distribution while keeping \tmax/\tmin\ fixed will only slide this
intermediate region towards higher or lower frequencies without changing the
\numax/\numin\ ratio. Although all spectral indices between $2$ and $-0.1$
occur in this intermediate range (see fig \ref{fi:hiispec}b), there exists a
`plateau' region where $\alpha$ changes relatively slowly. Integration over the
$\tau_0$ distribution (eq.\ \ref{eq:Nff}) shows that within this plateau
region, to leading order $f(\nu)$ varies as $\nu^{-2.1(\gamma - 1)}$ so that
the spectral index there is

\eq{\label{eq:alpha}
    \alpha \simeq 2 - 2.1(\gamma - 1)
}
The choice of $\gamma = 1.5$ therefore gives $\alpha \approx 0.95$.

Universality of the spectral shape holds only so long as $\N < 1$. At larger
cloud numbers the spectral shape becomes dependent on \N, with the
low-frequency $\nu^2$-behaviour approaching the high-frequency $\nu^{-0.1}$
domain as \N\ increases. From eq.\ \ref{eq:hiispec2}, the  $\nu^2$ behaviour
occurs whenever  $\tauE_\nu = \N f(\nu) >1$, which for $\N > 1$ is a condition
that holds up to the frequency at which $f(\nu) > 1/\N$ is still valid. Given
that in the intermediate domain  $f(\nu) \propto (\numin/\nu)^{-2.1(\gamma -
1)}$, this means that the lower boundary of the intermediate spectral index
range increases from \numin\ to $\numin\N^{1/2.1(\gamma - 1)}$, i.e. $\numin
\N^{0.95}$ when $\gamma = 1.5$.  Consistent with this predicted behaviour,
Figure~\ref{fi:hiispec}a shows  that as \N\ increases to become $\gg 1$, the
ensemble spectra saturate against the $\nu^2$ asymptote at increasingly higher
frequencies, a consequence of the fact that as  \N\  increases the population
has $\tauE_\nu=1$ up to higher frequencies. If we now consider the frequency
region at which the transition to an optically thin overall spectral index
$\alpha = -0.1$ occurs, then for all the values of $\N$ shown in
Figure~\ref{fi:hiispec}a this occurs at a frequency much larger than the one at
which $\tauE_\nu=1$  and hence  $\tauE_\nu  \ll 1$ at this upper turnover.  It
follows from eq.\ \ref{eq:hiispec2} that around the upper turnover the ensemble
spectrum remains proportional to  $\nu^2  f(\nu)$ scaled by $\N$ even when $\N
\gg 1$. The upper frequency of the intermediate spectral index region therefore
still occurs at the upper turnover frequency of $\nu^2  f(\nu)$, i.e. at the
frequency in which all clouds in the opacity distribution become optically thin
at $\numax = \nu_0 \tau_{\rm 0,max}^{1/2.1}$, independent of $\N$. As a
consequence of the different dependence  on  $\N$ of the lower and upper limits
of the intermediate spectral index range, the fractional frequency range
showing  intermediate spectral index shrinks with \N\ when $\N > 1$.
Specifically, when $\gamma = 1.5$, the fractional frequency range showing
intermediate spectral index approximates to $\N^{-0.95}(\tmax/\tmin)^{0.48}$;
this also explains the steepening of the spectral index  transition region with
increasing $\N$, evident in fig. \ref{fi:hiispec}(b).

Our results show that \cite{IGNACE04} correctly identified the unique set of
conditions in which clumpy HII regions with a power-law $\tau_0^{-\gamma}$
distribution of cloud optical depths can produce the observed $\alpha \simeq 1$
spectra. When $\N <1$, the relative width of the intermediate spectral index
region can be substantial when the clump $\tau_0$ distribution covers a large
range (eq.\ \ref{eq:range}). The observed spectral index in this intermediate
frequency range is controlled by $\gamma$, with $\gamma=1.5$ giving $\alpha
\simeq 1$ (eq.\ \ref{eq:alpha}).  Our formalism shows for the first time how
the output spectrum is affected when $\N > 1$, demonstrating that for such
cases the frequency range showing intermediate spectral index narrows
significantly with increasing $\N$.


\begin{figure*}[t!]
\centering
\includegraphics[width=\hsize]{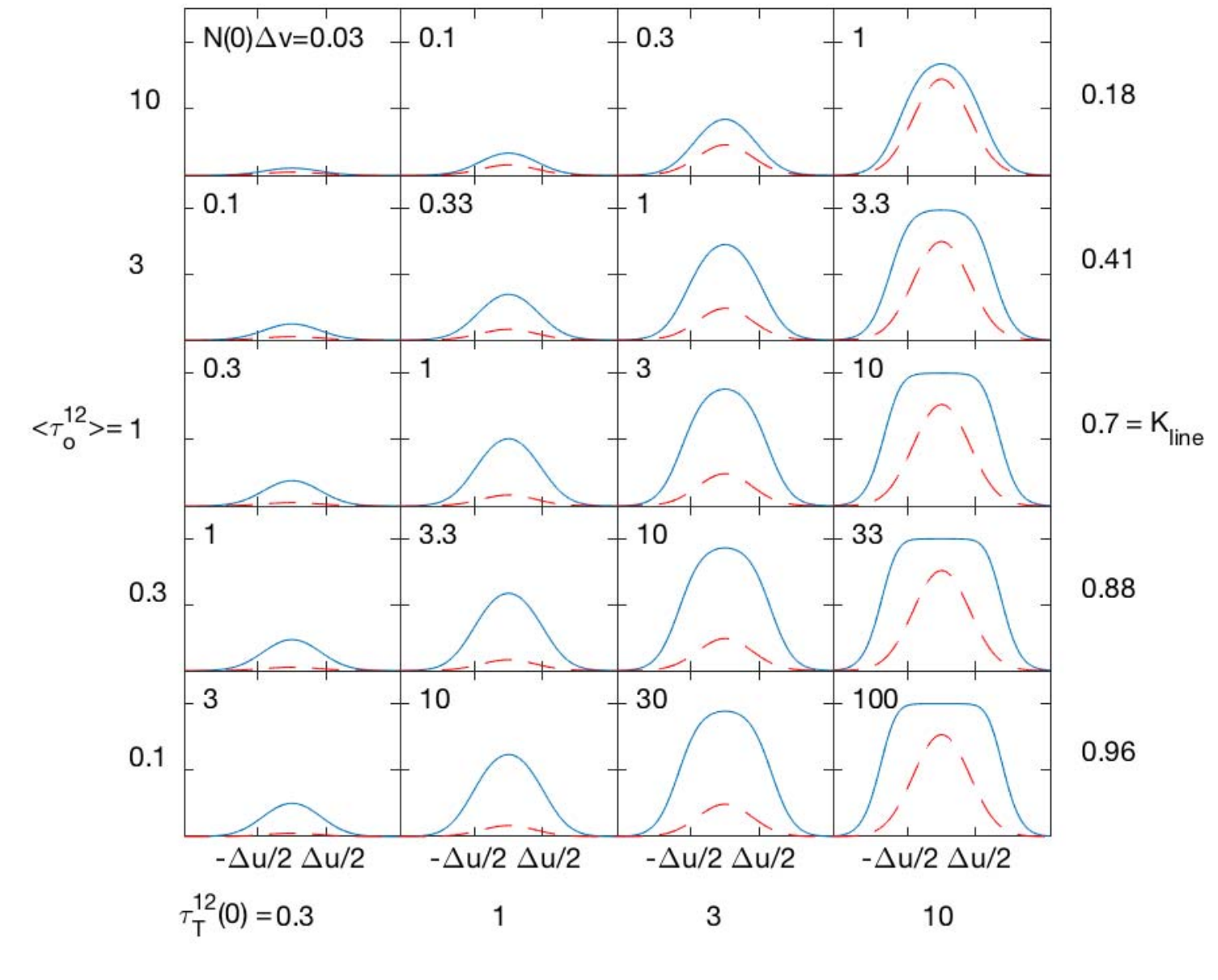}
\caption{Spectral line emission from an ensemble of clouds whose Gaussian LOS
velocity distribution width is $\Du \gg \Dv$, the internal velocity width of
each cloud. Shown is a sample of CO spectral profiles for different
combinations of individual {\it cloud}  optical depths (rows) and {\it total}
optical depths (columns).  Each panel shows plots of brightness temperature
versus velocity for \twelvCO (1--0) (solid blue lines) and \thirtCO (1--0)
(dashed red lines, multiplied by 5 to become more visible). Each $x$-axis is
centered on the line center, with tick marks shown at $\pm\frac12\Du$. Along
each $y$-axis, the top tick is placed at the \twelvCO\ line-center brightness
temperature of the bottom-right panel. Rows are labelled on the left side by
$\<\tau_{0}^{12}>$, the single cloud line-center \twelvCO\ optical depth
averaged over the cloud face, and on the right side by the corresponding line
clumping factor \Kline\ (eq.\ \ref{eq:kspec}). Columns are labeled at the
bottom by $\tauT^{12}(0)$, the total \twelvCO\ optical depth at line center
(eq.\ \ref{eq:tauT_line}). For $\tauT^{12}(0)$ to be the same for all the
panels in each column, the number of clouds intercepted, on average, along the
LOS within $\pm\frac12\Dv$ from line center must vary; this value is indicated
at the top left corner of every panel. For more details and interpretation see
text (\S\ref{se:line_profiles}).
}\label{fi:chess}
\end{figure*}


\subsection{CO Spectral Line Profiles}
\label{se:line_profiles}

Consider line emission from a population of small clouds within a telescope
beam coming from an external galaxy or a galactic star-forming region. The
resulting output spectrum is the  average of  the emission over many lines of
sight and therefore equals the LOS expectation value of spectral line emission
(see \S\ref{se:specemiss}). As an example of line emission from a clumpy
medium, Figure \ref{fi:chess} shows model CO spectra for ensembles of identical
uniform spherical clouds\footnote{As noted in Section \ref{se:shapes} the exact
cloud shape has very little effect on the emerging spectra from a cloud
population; for identical clouds, any cloud shape which has the same mean
opacity averaged over its face at cloud central velocity will have almost
identical spectra.}. The distribution of cloud LOS velocities is taken as
Gaussian with width \Du, assumed to be much larger than \Dv, the velocity width
of each cloud (as defined in \S\ref{se:specabs}). The expected spectrum in such
a case is given by eq.\ \ref{eq:simpspec}, which shows that it depends on the
product of the total optical depth \tauT(\v) and the clumping factor \Kline,
determined by the mean properties of clouds in the distribution (eq.\
\ref{eq:kspec}). Figure \ref{fi:chess} shows predicted \twelvCO(1--0) and
\thirtCO(1--0) spectra, with a \thirtCO(1--0) absorption coefficient that is
1/60 that of \twelvCO(1-0). Different columns show spectra for different values
of the total \twelvCO(1--0)  optical depth at zero velocity, $\tauT^{12}(0)$,
as listed at the bottom of each column.  Different rows present different
values of the mean single cloud $\<\tau_{0}^{12}>$, the line-center optical
depth averaged over the face of a cloud; the corresponding \Kline\ is listed to
the right of each row. Since $\tauT^{12}(0) = N(0) \<\tau_{0}^{12}> \Dv$ then
for each panel's combination of $\tauT^{12}(0)$ and $ \<\tau_{0}^{12}> $ there
is a corresponding value of $N(0)\Dv$, the number of clumps per LOS with center
velocity within $\pm \Dv/2 $ of zero velocity, as given in the top left corner
of each panel.

To understand the displayed profiles it helps to recall that the brightness
temperature of a smooth-density source with constant excitation temperature is
$\Tb = \Tx[1 - \exp(-\tau(\v))]$. Therefore, when the line-center optical depth
obeys $\tau(0) < 1$ the emission is proportional to $\tau(\v)$, displaying the
Gaussian velocity distribution profile. In contrast, when $\tau(0) > 1$ the
emission develops a flat-top profile that broadens with increasing $\tau(0)$.
This same familiar behavior is expected in a clumpy source when individual
clouds are optically thin, since clumping has no effect in that case
(\S\ref{se:conttrans}), with the profile dominated by the cloud ensemble
Gaussian motions since we assume $\Du \gg \Dv$. That is indeed what figure
\ref{fi:chess} shows. In all the panels, individual clouds are optically thin
in the \thirtCO(1-0) transition because of the \thirtCO\ small abundance. As a
result, the \thirtCO(1-0) line profiles are identical within the panels of each
column, for which total optical depth is the same, and every row shows the same
familiar evolution for an optically thin line profile---Gaussian with peak
intensity proportional to the line-center optical depth. In the bottom two rows
individual clouds are optically thin also in \twelvCO(1-0), and so in turn
display the same expected profile development with increasing total optical
depth to a fully saturated flat top in the rightmost panels. However, when
individual clouds become optically thick, clumping induces profile behavior
completely unexpected from the smooth-density experience. This unique behavior
stands out when panels are followed up the figure's rightmost column, where
$\tauT^{12}(0) = 10$: the flat top gets narrower as the optical depth per cloud
increases, until finally the original Gaussian profile emerges almost intact
when the line-center optical depth of a cloud is $\tau^{12}_0 = 10$. While it
is impossible in the case of a smooth-density gas that such a high $\tauT$
could have a velocity profile close to Gaussian, this can occur as a natural
consequence of clumpiness. Since the effective optical depth is $\tauE =
\Kline\tauT$ (eq. \ref{eq:kspec}), the line becomes effectively thin when the
optical depth of individual clouds is sufficiently high, a result of the
decline of \Kline\ (see fig.\ \ref{fi:kfactor}; note also the value of \Kline\
listed to the tight of each row). While in the top right panel of Figure
\ref{fi:chess} the overall optical depth \tauT\ is high, on average it is
concentrated into only one very optically thick cloud. Since the probability
distribution of clouds per LOS is a Poisson distribution, the fraction $e^{-1}$
(\about40\%) of all sightlines do not intercept any cloud. With only a fraction
of the surface area contributing to the observed radiation, the source
brightness temperature is less than the cloud excitation temperature and the
line remains unsaturated even at its center. At higher and lower velocities,
the covering factor of the optically thick clouds decreases and so the
resulting spectral profile closely follows the Gaussian profile of the number
of clouds per cloud LOS velocity width.

The richness of line spectra from clumpy media, indicated by the systematic
overview in Figure \ref{fi:chess} of profile variation with clump and total
optical depth, has important consequences for spectral studies of external
galaxies. It is important to note that although the calculations behind this
figure assume  the specific case of identical clouds, the resulting spectral
profiles are in fact universally applicable to a medium containing {\it any}
distribution of different cloud types. Equation \ref{eq:simpspec} shows that
clumpy medium emission is fully specified by \tauT\ and the clumping factor
\Kline, which is listed on the right of each row in the figure. A remarkable
result is that \emph{any medium containing a mixture of non-identical clouds
with the same value of \Kline\ as one of the rows of Fig \ref{fi:chess} will
have identical spectral shapes as a function of $\tauT^{12}(0)$, the total
optical depth at profile center.}

\subsubsection{Clumping Effects on Spectral Line Ratios}
\label{se:COlineratio}

Measurements of intensity ratios between spectral lines provide  a powerful
tool to constrain  chemical and physical conditions in astrophysical gases.
Such ratios depend on physical parameters such as the relative abundance of
molecular species and the excitation of the two transition lines. Often models
that fit line ratios assume uniform slabs of gas and neglect the potential
effects of clumping on small scales within these slabs.  However, if these
small-scale clumps do exist and are optically thick then clumping can
significantly impact observed line ratios; without accounting for such effects,
erroneous physical conditions can be deduced. Equation \ref{eq:simpspec} gives
the general expression for the observed line profile in the case of clumping,
with the profile intensity and shape depending on the brightness temperature of
the line, the total optical depth through the gas and the clumping factor,
$\Kline$; the latter, in turn, depends on the optical depth of individual
clouds (see eq.\ \ref{eq:kspec}).

Line ratios can be calculated employing either brightness profiles integrated
over all velocities of the two transitions or peak brightness temperatures at
the centers of the two profiles. Whichever procedure is used, there are in
general four separate opacity-related quantities that affect an observed line
ratio, comprising the total and single-cloud optical depths for each of the two
transitions. The line profiles shown in figure \ref{fi:chess} illustrate a
two-dimensional subspace of this general case in which the  total and
single-cloud optical depth of the weaker line (here \thirtCO(1-0)) are set $\ll
1$ while for the stronger line (\twelvCO (1-0)) these two quantities both vary
from $\ll 1$ to $\gg 1$.

The four panels in the bottom left corner of fig.\ \ref{fi:chess} illustrate
the situation when both total and single-cloud optical depths of
\mbox{\twelvCO(1-0)} are $\ll 1$, in which case there are no opacity effects.
In this regime the observed line ratio equals the intrinsic ratio of line
strengths determined by the gas physical conditions; for the displayed pair
this  is largely determined by the \twelvCO:\thirtCO\ abundance ratio. The four
panels in the figure bottom right corner have total $\tauT \gg 1$ but
individual clouds have $\tau \ll 1$. In this region the \twelvCO (1-0) line
profile saturates at line center, reducing the observed line ratio. In
principle, the saturation of the brighter line can be detected from the
relative line shapes of the two transitions, and so opacity effects on  the
line ratio can be corrected for. In practice, though, such corrections require
high signal-to-noise observations that are often not available.

The case of low total but high single-cloud optical depth is shown in the four
top left panels of fig.\ \ref{fi:chess}. In these spectra the high single-cloud
opacity causes clumping factors $\Kline <  1$, reducing the \twelvCO (1-0) line
effective optical depth (eq.\ \ref{eq:simpspec}) and hence its intensity, thus
reducing the line ratio by a factor of \Kline. Since the \twelvCO(1-0) line
shape is unaffected by having optically thick clouds, \emph{it is in principle
impossible to determine whether a reduction in observed line ratio reflects
clump opacity effects or an intrinsically low line ratio}. The final regime is
illustrated by the figure's four top right panels, where both single-cloud and
total optical depth are simultaneously high. Here both effects of reduction of
effective opacity and saturation of \twelvCO\ line profile are in operation;
their joint effect can be determined from eq.\ \ref{eq:simpspec}.

In the above we have carefully defined and distinguished the independent
quantities of total and single-cloud optical depth. Figure \ref{fi:chess}
illustrates that if {\it either}  of these optical depths is $>1$, in either
one of the transitions, then the observed line ratio will be significantly
different from its intrinsic value. Since high optical depths, either
single-cloud or total, often induce only small or imperceptible changes in line
shapes, it is usually not possible to detect their effect observationally. The
best that can be done is if a given transition is suspected (either {\it  a
priori} or based on other line ratios) to have {\it either} a high single-cloud
or total optical depth, then line ratios involving that transition should not
be used in fitting for gas physical conditions; such a transition could
reliably be used only to set limits on line ratios, which could then be
exploited in physical modelling.

\section{Discussion}
\label{se:discussion}

This paper brings to completion the \cite{NP84} approach to radiative transfer
in clumpy media, showing that such media can be reliably modeled as collections
of structureless clouds (``mega-particles'') characterized by a single
property---optical depth. The actual clouds can have a wide range of
properties, including different geometrical shapes, opacities, emissivities,
spectral shapes, bulk velocities, internal structures and orientations, all of
which can vary along the line of sight. With proper averaging, all of these
properties can be rigorously encapsulated in an ensemble of identical clouds,
and to a good degree of approximation, the geometry of these average clouds is
irrelevant.\footnote{\added{It is interesting to note the similarity with
approaches taken in the context of radiation propagation in porous media
\citep[see][]{Taine08}.}}

The simplicity of the formalism presented here has enabled us to readily
calculate clumpy emission spectra for a number of current problems, including
ensembles of ultra-compact HII regions (\S\ref{se:UCHII}), CO spectral lines
(\S\ref{se:line_profiles}) and synchrotron emission accompanied by free-free
absorption from supernovae and compact star-bursts in ultra-luminous IR
galaxies (Appendix \ref{se:ffabs}). Our results replicate and extend numerous
earlier studies. We show that \cite{IGNACE04} correctly identified the unique
set of conditions in which clumpy HII regions with a power-law distribution of
cloud optical depths can produce  $\alpha \simeq 1$ spectra (\S\ref{se:UCHII}).
And in the case of spectral line observations we show that it is impossible,
even in principle, to distinguish the effect of atomic and molecular abundances
on line ratios from the clumping effects of optically thick clouds
(\S\ref{se:COlineratio}).

The effective optical depth of a clumpy medium (eq.~\ref{eq:contabs}) arises
from the result of our formalism for $\<e^{-\tau_\nu}>$, the first moment of
the transmission factor distribution. Higher moments can be calculated just as
easily---the $m$-th moment, $e^{-m\tau_\nu}$, distribution average is simply
$\exp\left[-\N\left(1 - \left< e^{- m\tau_\nu}\right>\right)\right]$, as
directly obtained from the derivations in Appendix \ref{se_ap:contabs}. Such
moments can yield useful information about the cloud distribution.
\citet{TAUBER96} pointed out that a possible route to explore clumpiness is to
observe emission lines with very high spectral resolution and signal-to-noise
ratios, analyze the fluctuations in brightness temperature present on the line
shape, and infer from them the properties of the clumps present in the beam.
Assuming identical clumps and employing the \citet{MHS84} model,
\citet{TAUBER96}  computed the expected fluctuations for a wide range of clump
optical depths. Other studies of the fluctuations effect, both earlier
\citep{TAUBER91} and later \citep{PIROGOV12}, were restricted to optically thin
clumps. Based on our clump formalism it is straightforward to show that the
transmission-factor variance obeys
\begin{multline}
  \left\langle\left(e^{-\tau_\nu} - \<e^{-\tau_\nu}>\right)^2\right\rangle = \\
  \exp\left[ -\N\left(1 - \left< e^{-2\tau_\nu}\right>\right)\right] -
  \exp\left[-2\N\left(1 - \left< e^{- \tau_\nu}\right>\right)\right].
\end{multline}
This simple expression is completely general and encompasses the results of all
previous studies. It enables analysis of spectral variance, as performed by
\citet{TAUBER91} and \citet{TAUBER96}, for arbitrary sources without any model
restrictions. Such analysis can yield directly \N, the total number of clouds
along the LOS. Similar utility exists for higher moments, which can be derived
just as easily.

While the formalism developed here is quite general, it does rest on some
fundamental assumptions.  The clump volume filling factor is assumed small
enough that departures from Poisson statistics are small.  In practice, Monte
Carlo simulations (see \S\ref{se:simulations}) show the formalism to give good
estimates up to quite large volume filling factors (\about 10\%). When larger
filling factors are desired, modeling would have to rely on Monte Carlo
simulations. Next, absorption and emission properties of single clouds are
assumed unaffected by the radiation generated by the clumpy medium. Relaxing
this assumption requires an iterative procedure that starts with initial cloud
properties, such as level populations or dust temperature, determined in the
absence of cloud emission. In subsequent steps, the clump formalism is used to
calculate the expected radiation field, which is then added to the calculation
of individual cloud properties and reiterated until convergence.

Finally, the formalism assumes random placement in space of individual clouds,
such that the presence of another cloud nearby neither increases nor decreases
the probability for a cloud at a given position. However, there is evidence
suggesting that galactic clouds could be fractal \citep{FALG91, ELM04}, with
small high-density clumps embedded within larger lower-density clouds.
Nevertheless, there are reasons to expect our formalism to give approximately
correct answers even in such a case. Consider a critical size-scale at which
the optical depth is approximately unity, such that smaller, denser clumps are
optically thick but larger ones are optically thin. Then the fact that the
latter are correlated in position with clumps of the critical size has little
effect. Smaller, very optically thick cores will be embedded within the already
optically thick critical-scale clouds, thus their contributions to the total
spectrum will be small. Applying our formalism and considering only the clouds
at the critical size and larger should therefore produce reasonably accurate
results. We hope to quantify in future work the formalism accuracy when applied
to fractal clouds.

\appendix

\section{Analytic Results for Small Volume Filling Factors}
\label{se_ap:formalism}

\subsection{Cloud Distributions}
\label{se_ap:distributions}

The fundamental function describing the distribution of clouds with velocity
vector {\bf u} at position {\bf r} is $n({\bf r}, {\bf u})$, the number of
clouds per unit volumes of space and velocity space. When interacting with
continuum radiation, the cloud velocity is irrelevant and can be integrated out
to produce $n({\bf r}) = \int n({\bf r}, {\bf u})d^3{\bf u}$, the number of
clouds per unit volume at position {\bf r}. For radiation propagating along
some path, denote by $A$ the cloud area perpendicular to the path. Then the
number of clouds a photon encounters per unit length is $N = nA$, the inverse
of the local mean-free-path. As shown by \cite{NENKOVA02, NENKOVA08}, the
distribution $N$ suffices to describe the clumpy radiative transfer problem
when the volume filling factor is small; the volume density $n$ and the cloud
area $A$ do not enter separately in that case. Then the overall number of
clouds a photon encounters, on average, between any points $s_1$ and $s_2$
along the path is
\eq{
    \N(s_1,s_2) = \int_{s_1}^{s_2} N(s)ds
}
Generalizing to an arbitrary mix of cloud types is as simple as adding
independent variables, one for every additional cloud property, introducing the
distribution $N$ and deriving the corresponding \N\ for each type separately.

In the case of line radiation one needs to identify the clouds with a
particular LOS velocity $u$. To that end, integrate $n({\bf r}, {\bf u})$ over
the velocity components perpendicular to the path to get $n({\bf r},u) = \int
n({\bf r},{\bf u}) d\,^2u_\perp$, the cloud number density per unit volume and
unit LOS velocity $u$. As before, $N = nA$ and the number of clouds encountered
along the path between $s$ and $s + ds$ with LOS velocity between $u$ and $u +
du$ is $N(u,s)ds du$. Then the total number of clouds encountered between $s$
and $s + ds$ at all velocities is $N(s)ds$ while the total number encountered
along the entire path with velocities between $u$ and $u + du$ is $N(u)du$
where
\eq{\label{eq:N's}
    N(s) = \int N(u,s)du, \qquad
    N(u) = \int N(u,s)ds.
}
The total number of clouds along the entire path and at all LOS velocities is
\eq{
    \N = \int N(s)ds  = \int N(u)du
}

\subsection{Absorption in Clumpy Media}
\label{se_ap:contabs}

Consider the absorption of background radiation by a foreground ensemble of
identical clouds. Denote by $e^{-\tau_{\nu}}$ the transmission through a single
cloud at frequency $\nu$, and assume that interaction with the radiation does
not change $\tau_\nu$. Let $\N$ be the mean number of clouds along a given line
of sight. With the assumption that the clouds are placed randomly and their
volume filling factor is small, \citet{NP84} employed Poisson statistics to
show that the mean transmission factor averaged over many realizations of that
LOS is $\<e^{-\tau_{\nu}}> = \exp( -\tauE_\nu)$, where
\eq{\label{eq:basic}
   \tauE_\nu = \N(1 - e^{-\tau_{\nu}})
}
is the effective optical depth of the clumpy medium. Consider now a mixture of
$n$ types of clouds with each type having optical depth $\tau_{\nu,i}$ ($i=1,
2\ldots n $) and comprising a fraction $f_i = \N_{i}/\N$ of the average total
number of clouds along the LOS ($\sum f_i = 1$).  The probability of
encountering any given combination of number of clouds of each type is given by
a  multivariate Poisson probability distribution. Repeating the
\citeauthor{NP84} averaging procedure over all possible cloud types,
straightforward algebraic manipulations show that the effective optical depth
is now
\begin{equation}
\label{eq:aptau_av}
         \tauE_\nu = \N\left(1 - \<e^{-\tau_\nu}>\right)
\end{equation}
(cf eq.\ \ref{eq:contabs}), where
\eq{\label{eq:tau_av}
        \<e^{-\tau_\nu}> = \sum_i f_i \, e^{-\tau_{\nu,i}}
}
is the mean single-cloud transmission factor at frequency $\nu$. In the case
that the cloud type varies continuously with some parameter $t$, the discrete
fractional abundance $f_i$ is replaced by $f(t)dt$ (with $\int\!\! f(t)dt =
1$), the fraction of clouds in the parameter interval $[t, t + dt]$, and the
sum is replaced by the integral $\int\!\! f(t) e^{-\tau_{\nu}(t)} dt$. Note
that the variable $t$ can refer to any label defining cloud type. In the
simplest case it could refer to a single opacity characterizing each cloud and
$f(t)$ would then be the cloud opacity distribution.  It could also, however,
refer to the impact parameter of the LOS relative to the center of, say,
spherically symmetric clouds or an angle describing filamentary cloud
orientations (see Appendix \ref{se_ap:avg}). Furthermore, the cloud average can
be made over any number of dimensions of continuous parameters, $t_1, t_2
\ldots$, which describe the cloud population with the appropriate probability
distribution $f(t_1, t_2 \ldots)$.

The above results follow also from an alternative derivation that utilizes a
radiative transfer approach.\footnote{A somewhat similar approach has been used
by \cite{Lacki13}.} Start with the case of single-type clouds and introduce
$\eta(s) = N(s)/\N$, then the number of clouds encountered in differential
segment $ds$ along the radiation path is $N(s) = \N\eta(s) ds$ (note that
$\int\!\!\eta ds = 1$). Since each cloud absorbs the fractional amount $1 -
e^{-\tau_\nu}$ of impinging radiation, in traversing $ds$ the intensity is
attenuated by the amount $dI_\nu = -I_\nu\,\N (1 - e^{-\tau_\nu})\,\eta ds$.
Integrating along the full path yields the \citeauthor{NP84} result (eq.\
\ref{eq:basic}). Extending this approach to a cloud mixture is straightforward.
Introduce $\eta_i(s) = N_i(s)/\N_i$, the spatial distribution profile of
type-$i$ clouds ($\int\!\!\eta_i ds = 1$), then the number of such clouds
encountered in the differential segment $ds$ is $N_i(s) = \N f_i\eta_i(s) ds$.
Since each cloud absorbs the fraction $1 - e^{-\tau_{\nu,i}}$ of propagating
radiation, radiative transfer in clumpy media is controlled by
\eq{
    \frac{dI_\nu}{I_\nu} = -\N\sum_i f_i\left(1 - e^{-\tau_{\nu,i}} \right)
            \,\eta_i(s)ds
}
Integrating along the full path and summing over all cloud types leads directly
to $\tauE_\nu$ from eqs.\ \ref{eq:aptau_av} and \ref{eq:tau_av}.

\subsection{Emission from Clumpy Media}
\label{se_ap:contemiss}

The fundamental expression for continuum emission from single-type clouds has
been given in \citet{NENKOVA02, NENKOVA08}. To calculate the emission along a
given LOS in this case, denote by $\Sc(s)$ the single cloud source function at
position $s$; this is the increase in brightness of radiation propagating along
the LOS because of the emission from a single cloud. With $N(s)$ the expected
number of clouds per unit length, the overall number of clouds in differential
segment $ds$ along the path is $N(s)ds$ and the intensity generated in that
segment is $\Sc(s)N(s)ds$. This is the input radiation to the rest of the path,
which contains $\N(s) = \int_s N(s') ds'$ clouds on average. Therefore the mean
transmission factor for that remaining segment is $\exp[-\tauE_\nu(s)]$, where
$\tauE_\nu(s) = \N(s)(1 - e^{-\tau_\nu})$, and the emerging intensity is
\eq{
    I_{\rm C \nu} = \int e^{-\tau_{E\nu}(s)} \Sc(s)N(s)ds
}
Generalizing to a cloud distribution is straightforward. There are $N_i(s)$
clouds of type $i$  per unit length, on average, each with a source function
$\Sc_i(s)$. Then the intensity generated per unit length is $\sum_i N_i
\Sc_i(s) = N\<\Sc>$, where
\eq{\label{eq:Sav}
    \<\Sc> = \sum_i f_i\, \Sc_i
}
and a similar averaging procedure when the cloud distribution is characterized
by a continuous variable. Note that the clouds that dominate the averages in
$\<S>$ and in $\<e^{-\tau}>$ (eq.\ \ref{eq:tau_av}) can be different from each
other. Because the statistical variations of the emission and foreground
absorption are uncorrelated, the mean contribution of element $ds$ to the
received intensity equals its mean emission times the mean foreground
transmission. This leads directly to the result in eq.\ \ref{eq:contemiss}.

\subsection{Spectral Line Absorption}
\label{se_ap:specabs}

The optical depth of a line centered on frequency $\nu_0$ is $\tau(\nu) =
\tau_0\Phi(\nu - \nu_0)$, where $\tau_0$ is the line-center optical depth and
$\Phi$ is the line profile normalized to $\Phi(0) = 1$. The linewidth can be
defined from $\Dnu \equiv \int\Phi(\nu - \nu_0) d\nu$. In addition to the
potential intrinsic variation of $\tau_0$ and \Dnu\ from cloud to cloud,
$\nu_0$ can vary too because of cloud motions. The line-center frequency of a
cloud with LOS velocity $u$ is shifted to $\nu_0(1 - u/c)$ and its optical
depth becomes
\eq{
    \tau(\nu, u) = \tau_0\Phi[\nu - \nu_0(1 - u/c)]
}
Based on the approach described in \S\ref{se_ap:contabs} above, the effective
line opacity at frequency $\nu$ of an ensemble of clouds is given by
equation~\ref{eq:contabs} with the quantity $\<e^{-\tau_{\nu}}>$ now including
also an average over different cloud-center velocities.  Since we wish to
handle the cloud velocity distribution separately from all other possible
variations of cloud properties (i.e., $\tau_0$ and \Dnu), we lump them together
into a single symbolic variable $t$.  Then $\<e^{-\tau_{\nu}}> = \iint\!
f(u,t)e^{-\tau(\nu,u)}dudt$, where the distribution $f$ describes the
fractional number of clouds with LOS velocity $u$ and value $t$ for the other
cloud properties.

If the fraction of clouds of different types $t$ is independent of velocity
then $f(u,t) = g(t)N(u)/\N$, where \N\ is the overall number of clouds
encountered, on average, along the path, $N(u)$ is the number per unit LOS
bandwidth $u$ (see \S\ref{se_ap:distributions} above), and $g(t)$ is the
probability distribution of all other cloud properties ($\int g(t)dt = 1$).
Then equation \ref{eq:contabs} for the line effective optical depth becomes
\begin{equation}
  \tauE_\nu = \int N(u)
       \left(1 - \left\langle e^{-\tau(\nu,u)}\right\rangle\right) du,
\label{eq:specconv}
\end{equation}
where $\left\langle e^{-\tau(\nu,u)}\right\rangle = \int\!\! g(t)
e^{-\tau(\nu,u)}dt$ is the  cloud mean transmission factor, averaged over all
cloud properties other than LOS velocity. Instead of frequency $\nu$, the
equivalent Doppler velocity $\v = c(1 - \nu/\nu_0)$ is customarily employed.
Then the optical depth becomes $\tau(u,\v) = \tau_0\Phi(u - \v)$ and the
profile width is $\Dv = c\Dnu/\nu_0$, a velocity scale that characterizes the
cloud internal motions. Replacing frequency $\nu$ by equivalent Doppler
velocity \v\ in eq.\ \ref{eq:specconv} yields the result in eq.\
\ref{eq:tauE_line}.

\subsection{Spectral Line Emission}
\label{se_ap:spec_em}

Start with a population of identical clouds, described by the distribution
$N(s,u)$ (\S\ref{se_ap:distributions}). Denote by $S_0(s)$ the line center
brightness of a cloud at rest at position $s$, then the cloud emission at
frequency $\nu$ is $S_0\psi(\nu - \nu_0)$. When the cloud is optically thin
$\psi = \Phi$, the emission profile simply has the line (Doppler) shape. In the
general case, both $S_0$ and $\psi(\nu)$ are determined only after a detailed
solution of the radiative transfer problem for single clouds. When the cloud is
moving with Doppler velocity $u$, its emission at frequency $\nu$ becomes
$S_0(s)\psi[\nu - \nu_0(1 - u/c)]$, thus the average emission at frequency
$\nu$ from the cloud ensemble at position $s$ is described by the source
function
\eq{\label{eq:Psi}
    \Sc(s) = S_0(s)\Psi_\nu,
    \qquad \hbox{where} \quad
    \Psi_\nu = \frac{1}{N(s)}\int \psi[\nu - \nu_0(1 - u/c)]N(s,u)du
}
The intensity generated in segment $ds$ at frequency $\nu$ is $\Sc(s)N(s)ds$
and while travelling through the rest of the path it is attenuated by
$\exp[-\tauE_\nu(s)]$, where $\tauE_\nu(s)$, the line effective optical depth
from point $s$, is calculated from eq.\ \ref{eq:specconv}. Therefore, in
complete analogy with the continuum case (eq.\ \ref{eq:contemiss}) the emerging
line intensity is
\eq{\label{eq:line_S0}
    I_{\rm C\nu} = \int e^{-\tau_{E\nu}(s)}\Sc(s)N(s)ds
}
As before, generalizing this result to a mix of clouds is straightforward.
Adding an index $i$ to differentiate between cloud species and denoting by
$\Psi_{i\nu}$ the emission profile of the $i$-th species as defined in eq.\
\ref{eq:Psi}, the brightness generated per unit length is
\eq{\label{eq:line_S}
    \sum_i S_{0i}(s) \Psi_{i\nu}N_i(s)
        = N(s)\sum_i f_i(s) \Sc_i(s)(s)
        = N(s)\,\<\Sc(s)>
}
When the cloud mix is described by a continuous parameter, the sums in this
relation are trivially replaced by integrals. From the last two relations it
follows immediately that the line intensity emerging from a clumpy medium is
\eq{\label{eq:line_I}
    I_{\rm C \nu} = \int e^{-\tau_{E\nu}(s)}\<\Sc(s)>N(s)ds,
}
the same fundamental expression as for the continuum case (eq.\
\ref{eq:contemiss}).

Inside a cloud, the source function is $B_\nu(\Tx)$, where $B_\nu$ is the
Planck function and \Tx\ is the local line excitation temperature
\citep[e.g.,][]{MasersBook}. A widely used approximation is to assume a
constant \Tx, neglecting its variation with position inside the cloud. Then the
emission intensity of a single cloud at rest, the source function of the clumpy
medium, is $S_\nu = B_{\nu_0}(\Tx)\,\left(1 - e^{-\tau_\nu}\right)$. When the
velocity distribution is independent of position along the LOS this implies
$\Psi_\nu = \<1 - e^{-\tau_\nu}>$ as well as $\tauE_\nu = \N(s)\<1 -
e^{-\tau_\nu}>$. Expressing intensity in terms of equivalent brightness
temperature \Tb\ and assuming both \Tb\ and \Tx\ to be in the Rayleigh-Jeans
domain, eq.\ \ref{eq:line_I} yields
\eq{\label{eq:Tb}
   \Tb_\nu = \<\Tx> \left(1 - \<e^{-\tau_{E\nu}}>\right)
}
where we have additionally assumed that the locally-averaged excitation
temperature $\<\Tx>$ is the same everywhere in the clumpy region.

\section{Averages for Spheres, Slabs and Filaments}
\label{se_ap:avg}

The clumping correction factor involves the averages $\<\tau>$ and
$\<e^{-\tau}>$ over all cloud orientations (eq.\ \ref{eq:K}). In the case of
uniform clouds, these averages become straight geometrical integrations, which
we now consider for spherical and filamentary clouds. Additionally, thanks to
the symmetry properties of the planar geometry, the slab radiative transfer
problem is altogether independent of the density profile and is considered here
too. The frequency index is omitted since only the geometrical averaging is
considered here.

A uniform spherical cloud is fully characterized by its radius $R$ and optical
depth across the diameter $\tau_1$; the optical depth across a chord with
length $\ell$ is $\tau_1\,\ell/2R$. The pathlength along impact parameter $b$
is $\ell(b) = 2\sqrt{R^2 - b^2}$. Averaging over all impact parameters yields
\begin{subequations}\label{eq:sphere}
\begin{align}
    \<\tau>& = 2\tau_1\int_{0}^{1}\sqrt{1 - \rho^2}\, \rho d\rho
             = \frac23 \tau_1                             \label{eq:tau_sph} \\
    \<e^{-\tau}>&= 2\int_{0}^{1}\exp\left(-\tau_1\sqrt{1 - \rho^2}\right)
                                \rho d\rho
                = \frac{2}{\tau_1^2}\left[1 - (1 + \tau_1)e^{-\tau_1}\right]
                                                           \label{eq:exp_sph}
\end{align}
\end{subequations}
The analytic result for $\<e^{-\tau}>$ for uniform spheres was noted previously
by \citet{IGNACE04}.

\begin{figure}[!ht]
  \centering
 \includegraphics[width=0.7\hsize, clip]{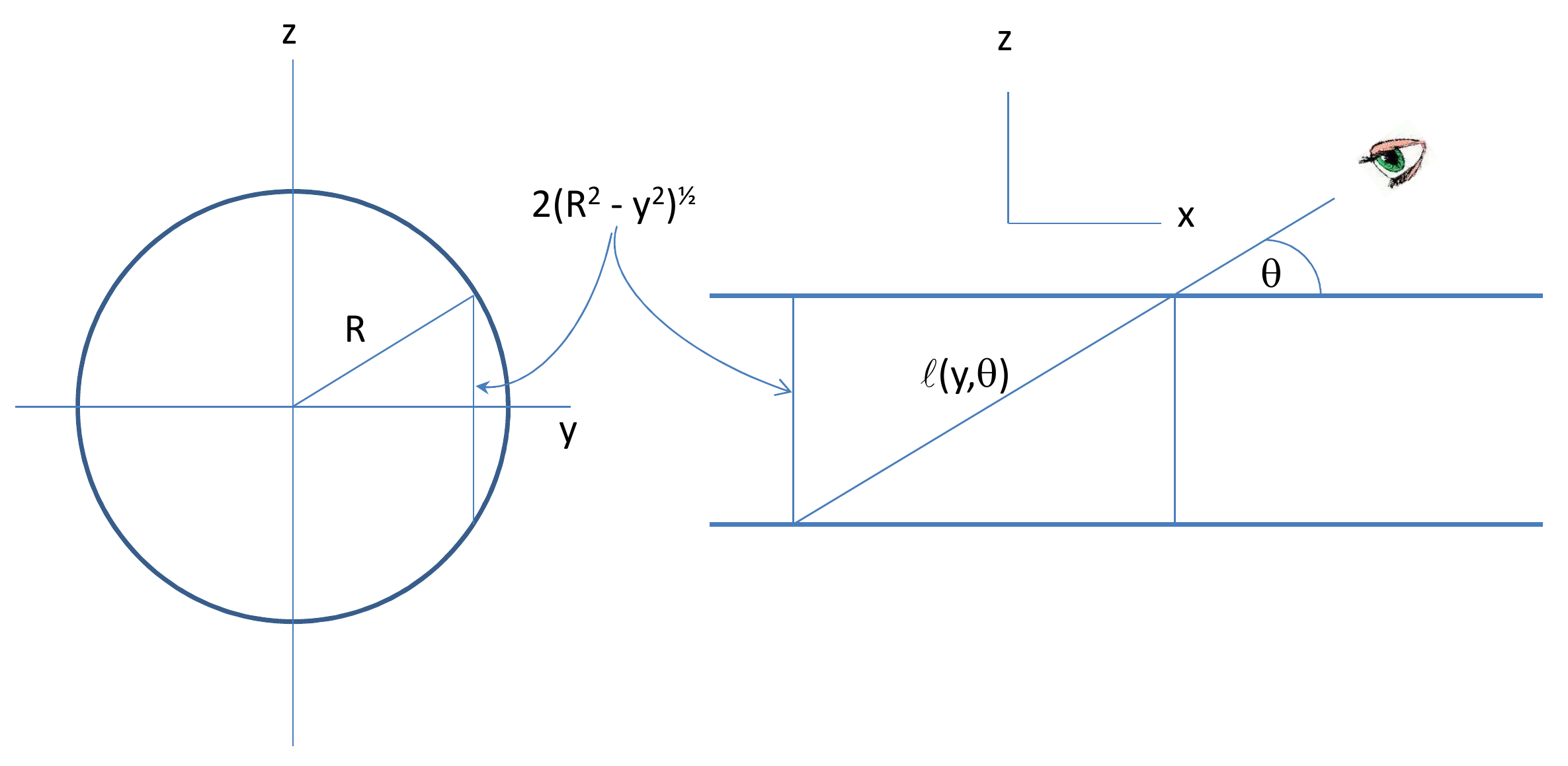}

\caption{A filament viewed at angle $\theta$ along a path displaced by impact
parameter $y$ from the axis. The cut on the left is perpendicular to the axis,
the one on the right is parallel to the axis and contains the viewing path.
}
  \label{fig:filament}
\end{figure}

Consider now a filament with radius $R$ and optical depth $\tau_1$ across the
diameter. The optical depth along a path displaced by distance $y$ and inclined
by angle $\theta$ from the axis (fig.\ \ref{fig:filament}) is
\eq{
   \tau(y,\theta) = \frac{\tau_1}{\sin\theta}\sqrt{1 - \rho^2},
   \qquad \hbox{where} \quad
   \rho = \frac{y}{R}
}
Significantly, $\tau$ does not depend on the filament's length---it depends
only on radius and viewing angle so long as the path does not intersect either
end cap. Assuming a large aspect ratio (length-to-width) so that the
contributions of the end caps can be neglected, averaging over $y$ and $\theta$
yields
\begin{subequations}
\begin{align}
    \<\tau>& = \tau_1\int_{0}^{1}d\mu \int_{0}^{1}
               \sqrt{\frac{1 - \rho^2}{1 - \mu^2}}d\rho
             = \frac{\pi^2}{8}\tau_1                       \label{eq:filament}\\
    \<e^{-\tau}>&= \int_{0}^{1} d\mu \int_{0}^{1}
          \exp\left(-\tau_1\sqrt{\frac{1 - \rho^2}{1 - \mu^2}}\right) d\rho
\end{align}
\end{subequations}
The planar geometry, with $\tau_1$ the optical depth along the normal, is
handled similarly. The pathlength is now independent of impact parameter while
having the same dependence on viewing angle, $\tau(\theta) =
\tau_1/\sin\theta$, leading to
\begin{subequations}\label{eq:slab}
\begin{align}
    \<\tau>& = \tau_1\int_{0}^{1}\frac{d\mu}{\sqrt{1 - \mu^2}}
             = \frac{\pi}{2}\tau_1                      \\
    \<e^{-\tau}>&= \int_{0}^{1}
          \exp\left(-\tau_1/\sqrt{1 - \mu^2}\right) d\mu
\end{align}
\end{subequations}
While the integrals for $\<e^{-\tau}>$ cannot be performed analytically for
slabs and filaments, the similarity with the spherical case of these integrals
explains the similarity of the $K$-factor as function of $\<\tau>$ for the
three geometries (fig.\ \ref{fi:kfactor}).

It may be worthwhile to note a simple result for the averaging over uniform
clouds of arbitrary shape.\footnote{This approach was first noted by Frank
Heymann.} With the $z$-axis along the line of sight and $x$-$y$ the plane of
the sky, the length through the cloud at point $(x,y)$ is $\ell(x,y)$.
Averaging over the observed area yields
\eq{\label{eq:lbar}
   \lbar = \frac{\int \ell(x,y) dxdy}{\int dxdy}
         = \frac{V}{\Ap}
}
where $V$ is the volume of the cloud and \Ap\ is its projected area on the
plane of the sky. With an arbitrary cloud shape we must further average over
all cloud orientations. Denote such averaging by $\<>$ then the desired
quantity is
\eq{\label{eq:lavg}
    \lavg = V\left\langle \frac{1}{\Ap} \right\rangle
}
The results for $\<\tau>$ for both uniform spheres (eq.\ \ref{eq:tau_sph}) and
filaments (eq.\ \ref{eq:filament}) are readily recovered from this general
expression.

\section{Synchrotron Emission with Free-Free Absorption}
\label{se:ffabs}


\begin{figure*}[t!]
\centering
\includegraphics[width=0.8\hsize]{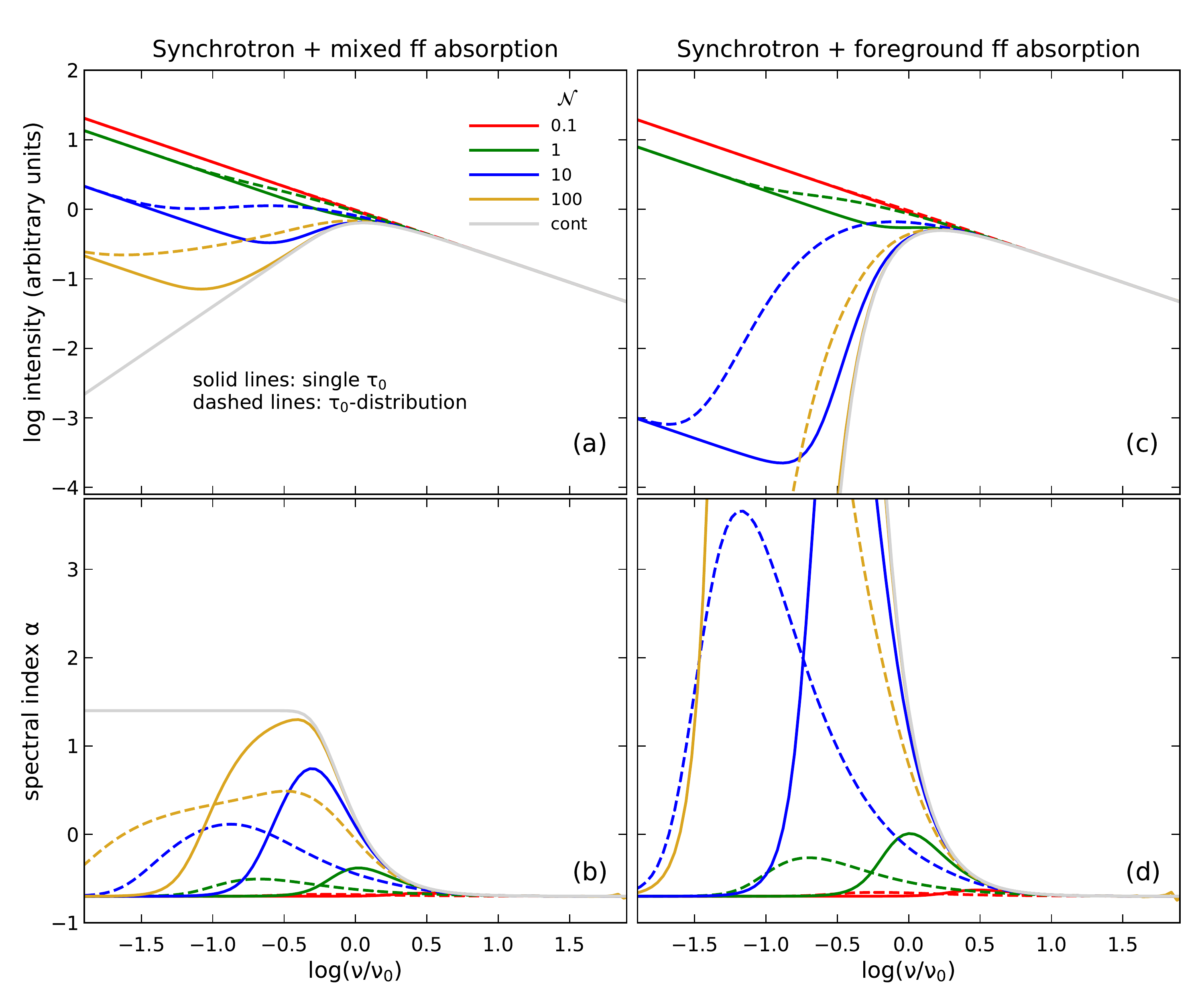}

\caption{Model spectra from synchrotron emission in the presence of free-free
absorption. All models have the same value of mean LOS free-free Emission
Measure through the source but spectra calculated for different geometries and
degrees of clumping. The left panels are for a geometry in which the
synchrotron emission and free-free absorption are intermixed, with panel
(\emph{a}) showing the resulting spectra and (\emph{b}) the spectral indices.
The right panels (\emph{c}) and (\emph{d}) show respectively spectra and
spectral indices when the free-free absorption is in a screen foreground to the
synchrotron emission. For both geometries  the  synchrotron emission is assumed
to follow $S_{\nu} = (\nu^/\nu_0)^{-0.7}$, with $\nu_0$ the fiducial frequency
defined such that for all models the mean {\it total} free-free optical depth
through the source is $\tauT(\nu_0) = 1$.  In all panels, the grey curves
(labeled cont) show the case that the free-free absorbing gas has a smooth
distribution, colored lines instead show models in which this gas is clumpy
with a varying mean number of clouds per LOS, \N, as labeled.  Solid colored
lines show cases with identical clumps all having the same value of optical
depth at the fiducial frequency, $\tau_0=1/\N$.  Dashed colored lines are in
contrast for models with cloud populations containing a wide distribution of
$\tau_0$ (see text for details). In all panels, the \N\ = 0.1 curves, which are
almost identical for the solid and dashed lines, are applicable to all $\N < 1$
models. Note that in panels (c) and (d), the \N\ = 100 solid-line curves are
virtually indistinguishable from the smooth-density absorption case.}
 \label{fi:ffabs2}
\end{figure*}


Observations of radio emission from compact star-bursts in ultra-luminous IR
galaxies are commonly fitted by  a model first proposed by \citet{CONDON91}
that combines the effects of star-formation induced synchrotron emission and
free-free absorption. This specific model belongs to a general class of
two-component mixtures of uniformly distributed emitters and absorbers;
examples include ionized gas mixed with dust \citep{NP84} and a uniform mixture
of stars and dust \citep{Thronson90}. In these admixtures, radiation generated
by the emitting component (ionized gas, stars, etc.) is selectively attenuated
by the smoothly distributed (i.e., not clumped) absorbing material, which
itself does not emit appreciably at the relevant wavelengths. Consider a slab
containing such a mixture with $\tauT_{\nu}$ the total optical depth for the
absorbing component and denote by $S_\nu$ the intensity of emission that would
have emerged in the absence of the absorbers. Assuming the emitters and
absorbers to be well-mixed together, the emission is everywhere proportional to
the absorption so that the intensity generated between $\tau_\nu$ and $\tau_\nu
+ d\tau_\nu$ is $S_\nu d\tau_\nu/\tauT_{\nu}$. On its way out this radiation is
attenuated by $\exp(-\tau_{\nu})$, emerging as $dI_\nu = S_{\nu}e^{-\tau_{\nu}}
d\tau_\nu/\tauT_{\nu}$. Integrating over the path, the emerging intensity is
now $I_\nu = S_{\nu}\T$, instead of $S_\nu$, where
\eq{\label{eq:Ts}
  \T  = \frac{1}{\tauT_{\nu}}\int e^{-\tau_{\nu}}d\tau_{\nu}
      = \frac{1 - \exp(-\tauT_{\nu})}{\tauT_{\nu}}
}
is the transmission factor for the smooth-density absorbing component. This
factor has the same functional form as the clumping factor $K_\nu$ (eq.\
\ref{eq:K}), approaching the limits $\T \simeq 1$ when $\tauT_{\nu}  < 1$ and
$\T \simeq1/\tauT_{\nu} $ when $\tauT_{\nu} > 1$; in the former case the
radiation emerges intact from the entire slab, in the latter it emerges only
from within \about 1 optical depth from the surface, and the fractional
thickness of this surface layer is $1/\tauT_{\nu}$.

In the \citet{CONDON91} model, synchrotron emission $S_{\nu} \propto \nu^{-p}$,
with $p \sim 0.7$ typically, is attenuated by free-free absorbing gas well
mixed together with the emitting gas.\footnote{A factor $1/\tau_{ff}$ is
missing from eq.\ 8 of \citet{CONDON91}.} Emission from the free-free gas can
be neglected at radio wavelengths so that the emergent radiation is just the
absorbed synchrotron emission $ I_\nu  = S_{\nu} \T$, where \T\ is the
free-free transmission factor from eq.\ \ref{eq:Ts}. Denote by $\nu_0$ the
frequency where the free-free optical depth is unity, then $\tauT_{\nu} =
(\nu_0/\nu)^{2.1}$ (see eq.\ \ref{eq:tauff}). At frequencies higher than
$\nu_0$, $\tauT_{\nu} < 1$ and the emergent intensity is $I_\nu \propto
\nu^{-p}$, while at lower frequencies $\tauT_{\nu}  > 1$ and $I_\nu \propto
\nu^{2.1 - p}$. Therefore the spectrum peaks at $\nu_0$, falling away towards
both lower and higher frequencies; this behavior is shown by the  grey line in
panel (a) of Figure \ref{fi:ffabs2}.

We propose an extension of the \citet{CONDON91} model in which the synchrotron
emission remains smoothly distributed but the free-free absorption is clumpy.
This would be the case, for instance, if the absorption came from individual
HII regions within a star-forming region (\citealt{Lacki13}) while the
synchrotron emission came from the inter-clump medium. As shown in
\S\ref{se:conttrans}, in this case the attenuation is controlled by the
effective optical depth \tauE\ (eq.\ \ref{eq:contabs}). Assume that the density
of clumps along the line of sight is always proportional to the synchrotron
emission (because both are proportional to the star-formation rate) and further
assume that the clumping factor is constant along the LOS. Given these
assumptions, exactly the same derivation as above for the smooth-absorption
case is applicable, with \tauE\ replacing \tauT\ everywhere. The resulting
emergent intensity again is $I_\nu = S_{\nu}\T$, where now
\eq{\label{eq:Tc}
      \T =   \frac{1 - \exp(-\tauE_{\nu})}{\tauE_{\nu}},
}
with $\tauE_{\nu}$ the overall effective optical depth (eq.\ \ref{eq:contabs}).
The smooth-absorption result in eq.\ (\ref{eq:Ts}) is recovered when
$\tauE_{\nu}$ is replaced by the total optical depth $\tauT_{\nu}$, therefore
our extended model contains the \citet{CONDON91} model as a limiting  case.
Because $\tauE_{\nu}= K_{\nu}\tauT_{\nu}$ (eq.\ \ref{eq:Kdef}), clumpiness
affects the emergent radiation only at frequencies where the clumping factor
$K_{\nu}$ deviates from unity, i.e., only at frequencies at  which  individual
clumps are optically thick. In that case $K_{\nu} < 1$ and the mean effective
optical depth is reduced because of the possibility that by chance, no
optically thick clumps are encountered by radiation propagating out of the
slab.

Consider now keeping the mean total LOS optical depth at the fiducial $\nu_0$
fixed at $\tauT_{\nu_0} = 1$, with all this absorption concentrated into a
varying number \N\ of identical clouds. This corresponds to a situation in
which the mean integrated free-free Emission Measure (EM in units
pc\,cm$^{-6}$) per LOS is kept constant but the  EM per clump varies. The
emergent spectra are shown with solid colored curves in panel (a) of Fig
\ref{fi:ffabs2}, while panel (b) shows the spectral index of each model. The
unique properties of clumpy absorption stand out immediately in the \N\ = 0.1
plot---the input synchrotron radiation emerges unperturbed at all frequencies,
even $\nu < \nu_0$ where $\tauT_{\nu} > 1$. The reason is that in crossing the
entire region, the radiation can avoid all clouds with a probability $e^{-\cal
N}$, which is $\simeq 1$ when $\N < 1$. Thus the \N\ = 0.1 curve in Fig
\ref{fi:ffabs2}a is virtually identical to the input power-law spectrum and is
representative of all $\N < 1$ models. When $\N > 1$, a fraction $1 - e^{-\cal
N}$ of the radiation will interact with the absorbing clouds, resulting in two
spectral regimes depending on the optical depth $\tau_\nu$ of individual
clouds. When $\tau_\nu < 1$, clumping is irrelevant and the emergent radiation
is the same as for smooth-density absorption, while when $\tau_\nu > 1$,
$\tauE_{\nu} = \N$ (eq.\ \ref{eq:contabs}), the overall transmission factor is
$\T \simeq 1/\N$ (eq. \ref{eq:Tc}) and the emergent intensity is $S_\nu/\N$.
The observed radiation switches from the spectral shape of the
smooth-absorption model to that of the input synchrotron, albeit at a reduced
amplitude, at the frequency $\nu_N$ that gives $\tau(\nu_N) = 1$ for each
clump, i.e., a total optical depth $\tauT_{\nu_N} = \N$. Since $\tauT_{\nu_0} =
1$, it follows that
\eq{\label{eq:nuN}
   \nu_{N} = \frac{\nu_0}{\N^{0.48}}\,.
}
This transition is evident in the displayed cases of \N\ = 10 and 100, which
have $\nu_{N}/\nu_0$ = 0.33 and 0.11  respectively---at $\nu > \nu_N$ the
solutions overlap with the smooth-density result, while at  $\nu < \nu_N$ the
output spectrum equals the input synchrotron at reduced strength. As \N\ is
further increased, $\nu_{N}$ is decreasing and the deviation from the
smooth-density absorption is moving to the left. The smooth-absorption case is
fully recovered in the formal limit $\N \to \infty$, where $\nu_{N} \to 0$ and
all clouds are optically thin at all frequencies.

While the above discussion assumes  that all the absorbing clumps are
identical, the formalism can readily deal with the more realistic scenario of a
wide distribution in clump properties. The dashed colored curves in Fig
\ref{fi:ffabs2}a show example spectra for such cases. As in the case of
ultracompact HII regions (\S\ref{se:UCHII}), the distribution of clump optical
depths at frequency $\nu_0$ follows eq.\ \ref{eq:Nff} with $\gamma = 1.5$ and
\tmax/\tmin\ = \E4. Different colored dashed lines again correspond to
different choices for \N, the mean number of clumps per LOS. In each such case,
\tmin\ of the clump opacity distribution is adjusted so that the mean total
optical depth at $\nu_0$ averaged over all clumps is  unity, i.e., for each
$\N$ the value of \tmin\ is adjusted so that, averaged over the
\mbox{$\tau_0$-distribution}, the mean single-clump optical depth at the
fiducial frequency $\nu_0$ is $\<\tau_0> = 1/\N$. Compared with the case of
identical clumps (solid lines), in the clump distribution case the frequency
range over which the spectral index is flatter than the pure synchrotron
spectrum is much wider while the peak positive spectral index is reduced in
value (see Fig \ref{fi:ffabs2}b); these changes occur because the relatively
sharp transition at $\nu = \nu_N$ (eq.\ \ref{eq:nuN}) for identical clumps is
now spread over a range of frequencies. It is interesting to note that the
flattened spectral shape of the \N\ = 10 case in Fig \ref{fi:ffabs2}a (blue
dashed line) and the corresponding spectral index in Fig \ref{fi:ffabs2}b are
similar to some recent low-frequency  observations of star-forming galaxies
\citep{MARVEL15, CALISTRO17, GALVIN18}. Clumpy free-free absorption with a wide
range of clump opacities mixed with a synchrotron emitting medium is thus a
possible explanation.\footnote{Alternative explanations include low frequency
modifications in the energy power law index of lower energy synchrotron
emitting electrons, or spatially separated source components, which are
free-free absorbed at different frequencies.}

As with the original \citet{CONDON91} work, all model spectra presented here
ignore contributions from free-free emission.  For typical star
formation-powered radio emission, where clump free-free absorption only becomes
significant below 1GHz, this is a good approximation up to 20--100 GHz. A more
complete, integrated spectrum can be calculated by combining the model clump
free-free emission spectrum from a population of clumps with the spectra
described in \S\ref{se:UCHII} (and shown in Fig \ref{fi:hiispec}). We plan to
present such overall spectra fitted to galaxy SEDs in a future publication
\vskip 0.7cm

\subsection{Absorption by Foreground Screen}
\label{se:fffore}

Observations of the evolving radio spectra of supernovae are broadly explained
by synchrotron emission from the supernova shell that passes through a
foreground free-free absorbing ionized wind from the progenitor star
\citep{WEILER86, WEILER02}. It is now thought that mass loss from such massive
stellar progenitors is clumped \citep{SMITH14}; indeed, observations of radio
supernovae show evolving radio spectra which are not always compatible with a
smooth progenitor wind \citep{VANDYK94, WEILER02}.

In calculating the intensity $I_\nu$ of background radiation that passes
through foreground absorption, the only difference from the internal-absorption
discussed above is that the functional form of the transmission factor \T\ is
replaced by an exponential with the same argument---$\tauT_\nu$ when the
absorbing material is distributed smoothly (cf. eq.\ \ref{eq:Ts}) and
$\tauE_\nu$ when it is clumpy (cf. eq.\ \ref{eq:Tc}). Therefore, in the case of
smooth-density absorption the emerging spectra differ in the two scenarios only
at $\nu < \nu_0$, where the foreground screen yields $I_\nu =
S_\nu\exp(-\tauT_\nu)$ instead of $I_\nu = S_\nu/\tauT_\nu$ for internal
obscuration and large $\tauT_\nu$. In the case of clumpy absorption with $\N <
1$, the input radiation passes through almost unmodified irrespective of the
geometry thanks to the high probability to avoid all clouds. And when $\N > 1$,
clumpy absorption deviates from its smooth-density counterpart only at $\nu <
\nu_N$ (eq.\ \ref{eq:nuN}), where the foreground screen yields  $I_\nu = S_\nu
e^{-\cal N}$ instead of $S_\nu/\N$ for internal obscuration and large $\N$.
Foreground screens can thus produce, for both smooth and clumpy absorption,
much steeper spectral falloffs at low frequency than internal obscuration.

Panel (c) of Figure \ref{fi:ffabs2} shows synchrotron spectra attenuated by
free-free absorbing foreground screens with both smooth and clumpy density
distributions, repeating all the parameter combinations of the previous
section; panel (d) shows the spectral index of each model. The above analysis
is borne out by the numerical calculations. In particular, the synchrotron
radiation emerges almost unmodified when \N\ = 0.1; this happens in all $\N \ll
1$ models, whatever the geometrical configuration. And compared with the
corresponding internal-absorption results, the spectral declines are much
steeper for both smooth-absorption at $\nu < \nu_0$ and $\N > 1$ clumpy
absorption at $\nu < \nu_N$. With the dynamic range covered in the figure,
clumpy foreground screens with $\N > 10$ (such as the displayed \N\ = 100) are
indistinguishable from smooth ones.  As is evident from panels (b) and (d), a
foreground screen can produce extreme cases of spectral index in comparison
with internal obscuration.

The problem of synchrotron emission with free-free absorption contains two
elements that combine into four possible configurations: the absorbing material
can be internal or external to the emission region, and its distribution can be
either smooth or clumpy. The literature on fitting the spectra and light curves
of radio supernovae \citep[e.g.,][]{WEILER02} is not always clear on the
distinctions between the different combinations and the origins of the factors
accounting for the various types of absorption. As shown here, smooth-density
absorption is described by the transmission factor $T(\tauT_\nu)$ (eq.\
\ref{eq:Ts}) when it is internal and by $\exp(-\tauT_\nu)$ when it is external.
When the absorbing material is clumped, the overall optical depth  $\tauT_\nu$
is simply replaced in either case by the effective optical depth  $\tauE_\nu$
(eq.\ \ref{eq:contabs}). And since $\tauE_\nu  = K_{\nu}\tauT_\nu$ (eq.\
\ref{eq:Kdef}), the only effect of clumpiness is to modify the optical depth by
a clumping correction factor $K_{\nu}$ (eq.\ \ref{eq:K}), whether or not the
extinction is internal or external.

\bigskip
\noindent\emph{Acknowledgements}: \added{We are grateful to the anonymous
referee for useful comments.} ME acknowledges the help of Frank Heymann and
Robert Nikutta.

\bigskip


\begin{thebibliography}{}
\expandafter\ifx\csname natexlab\endcsname\relax\def\natexlab#1{#1}\fi
\providecommand{\url}[1]{\href{#1}{#1}}
\providecommand{\dodoi}[1]{doi:~\href{http://doi.org/#1}{\nolinkurl{#1}}}
\providecommand{\doeprint}[1]{\href{http://ascl.net/#1}{\nolinkurl{http://ascl.net/#1}}}
\providecommand{\doarXiv}[1]{\href{https://arxiv.org/abs/#1}{\nolinkurl{https://arxiv.org/abs/#1}}}

\bibitem[{{Calistro Rivera} {et~al.}(2017){Calistro Rivera}, {Williams},
  {Hardcastle}, {Duncan}, {R{\"o}ttgering}, {Best}, {Br{\"u}ggen}, {Chy{\.z}y},
  {Conselice}, {de Gasperin}, {Engels}, {G{\"u}rkan}, {Intema}, {Jarvis},
  {Mahony}, {Miley}, {Morabito}, {Prandoni}, {Sabater}, {Smith}, {Tasse}, {van
  der Werf}, \& {White}}]{CALISTRO17}
{Calistro Rivera}, G., {Williams}, W.~L., {Hardcastle}, M.~J., {et~al.} 2017,
  \mnras, 469, 3468, \dodoi{10.1093/mnras/stx1040}

\bibitem[{{Condon} {et~al.}(1991){Condon}, {Huang}, {Yin}, \&
  {Thuan}}]{CONDON91}
{Condon}, J.~J., {Huang}, Z.-P., {Yin}, Q.~F., \& {Thuan}, T.~X. 1991, \apj,
  378, 65, \dodoi{10.1086/170407}

\bibitem[{{Elitzur}(1992)}]{MasersBook} {Elitzur}, M. 1992, Astronomical Masers
    (Kluwer Academic Publishers),
  \dodoi{10.1007/978-94-011-2394-5}

\bibitem[{{Elmegreen} \& {Scalo}(2004)}]{ELM04} {Elmegreen}, B.~G., \& {Scalo},
    J. 2004, \araa, 42, 211,
  \dodoi{10.1146/annurev.astro.41.011802.094859}

\bibitem[{{Falgarone} {et~al.}(1991){Falgarone}, {Phillips}, \&
  {Walker}}]{FALG91}
{Falgarone}, E., {Phillips}, T.~G., \& {Walker}, C.~K. 1991, \apj, 378, 186,
  \dodoi{10.1086/170419}

\bibitem[{{Galvin} {et~al.}(2018){Galvin}, {Seymour}, {Marvil},
  {Filipovi{\'c}}, {Tothill}, {McDermid}, {Hurley-Walker}, {Hancock},
  {Callingham}, {Cook}, {Norris}, {Bell}, {Dwarakanath}, {For}, {Gaensler},
  {Hindson}, {Johnston-Hollitt}, {Kapi{\'n}ska}, {Lenc}, {McKinley}, {Morgan},
  {Offringa}, {Procopio}, {Staveley-Smith}, {Wayth}, {Wu}, \&
  {Zheng}}]{GALVIN18}
{Galvin}, T.~J., {Seymour}, N., {Marvil}, J., {et~al.} 2018, \mnras, 474, 779,
  \dodoi{10.1093/mnras/stx2613}

\bibitem[{{Ignace} \& {Churchwell}(2004)}]{IGNACE04} {Ignace}, R., \&
    {Churchwell}, E. 2004, \apj, 610, 351, \dodoi{10.1086/421453}

\bibitem[{{Lacki}(2013)}]{Lacki13} {Lacki}, B.~C. 2013, \mnras, 431, 3003,
    \dodoi{10.1093/mnras/stt349}

\bibitem[{{Laor} {et~al.}(2006){Laor}, {Barth}, {Ho}, \& {Filippenko}}]{LAOR06}
    {Laor}, A., {Barth}, A.~J., {Ho}, L.~C., \& {Filippenko}, A.~V. 2006, \apj,
  636, 83, \dodoi{10.1086/497908}

\bibitem[{{Martin} {et~al.}(1984){Martin}, {Hills}, \& {Sanders}}]{MHS84}
    {Martin}, H.~M., {Hills}, R.~E., \& {Sanders}, D.~B. 1984, \mnras, 208, 35

\bibitem[{{Marvil} {et~al.}(2015){Marvil}, {Owen}, \& {Eilek}}]{MARVEL15}
    {Marvil}, J., {Owen}, F., \& {Eilek}, J. 2015, \aj, 149, 32,
  \dodoi{10.1088/0004-6256/149/1/32}

\bibitem[{{Natta} \& {Panagia}(1984)}]{NP84} {Natta}, A., \& {Panagia}, N.
    1984, \apj, 287, 228, \dodoi{10.1086/162681}

\bibitem[{{Nenkova} {et~al.}(2002){Nenkova}, {Ivezi{\'c}}, \&
  {Elitzur}}]{NENKOVA02}
{Nenkova}, M., {Ivezi{\'c}}, Z., \& {Elitzur}, M. 2002, \apjl, 570, L9,
  \dodoi{10.1086/340857}

\bibitem[{{Nenkova} {et~al.}(2008){Nenkova}, {Sirocky}, {Ivezi{\'c}}, \&
  {Elitzur}}]{NENKOVA08}
{Nenkova}, M., {Sirocky}, M.~M., {Ivezi{\'c}}, Z., \& {Elitzur}, M. 2008, \apj,
  685, 147, \dodoi{10.1086/590482}

\bibitem[{{Pirogov} {et~al.}(2012){Pirogov}, {Zinchenko}, {Johansson}, \&
  {Yang}}]{PIROGOV12}
{Pirogov}, L.~E., {Zinchenko}, I.~I., {Johansson}, L.~E.~B., \& {Yang}, J.
  2012, Astronomical and Astrophysical Transactions, 27, 475

\bibitem[{{Smith}(2014)}]{SMITH14} {Smith}, N. 2014, \araa, 52, 487,
    \dodoi{10.1146/annurev-astro-081913-040025}

\bibitem[{{Taine} {et~al.}(2008){Taine}, {Iacona}, \&
  {Bellet}}]{Taine08}
{{Taine}, J., {Iacona}, E., \& {Bellet}, F. 2008, in 5th European
Thermal-Sciences Conference (EUROTHERM), pp.ISBN 978-90-386-1274-4 $\rm\<
hal-00340096>$}


\bibitem[{{Tauber}(1996)}]{TAUBER96} {Tauber}, J.~A. 1996, \aap, 315, 591

\bibitem[{{Tauber} {et~al.}(1991){Tauber}, {Goldsmith}, \&
  {Dickman}}]{TAUBER91}
{Tauber}, J.~A., {Goldsmith}, P.~F., \& {Dickman}, R.~L. 1991, \apj, 375, 635,
  \dodoi{10.1086/170226}

\bibitem[{{Thronson} {et~al.}(1990){Thronson}, {Majewski}, {Descartes}, \&
  {Hereld}}]{Thronson90}
{Thronson}, Jr., H.~A., {Majewski}, S., {Descartes}, L., \& {Hereld}, M. 1990,
  \apj, 364, 456, \dodoi{10.1086/169430}

\bibitem[{{van Dyk} {et~al.}(1994){van Dyk}, {Weiler}, {Sramek}, {Rupen}, \&
  {Panagia}}]{VANDYK94}
{van Dyk}, S.~D., {Weiler}, K.~W., {Sramek}, R.~A., {Rupen}, M.~P., \&
  {Panagia}, N. 1994, \apjl, 432, L115, \dodoi{10.1086/187525}

\bibitem[{{Wall}(2006)}]{WALL06} {Wall}, W.~F. 2006, Revista Mexicana de
    Astronomia y Astrofisica, 42, 117

\bibitem[{{Wall}(2007)}]{WALL07}
---. 2007, \mnras, 379, 674, \dodoi{10.1111/j.1365-2966.2007.11968.x}

\bibitem[{{Weiler} {et~al.}(2002){Weiler}, {Panagia}, {Montes}, \&
  {Sramek}}]{WEILER02}
{Weiler}, K.~W., {Panagia}, N., {Montes}, M.~J., \& {Sramek}, R.~A. 2002,
  \araa, 40, 387, \dodoi{10.1146/annurev.astro.40.060401.093744}

\bibitem[{{Weiler} {et~al.}(1986){Weiler}, {Sramek}, {Panagia}, {van der
  Hulst}, \& {Salvati}}]{WEILER86}
{Weiler}, K.~W., {Sramek}, R.~A., {Panagia}, N., {van der Hulst}, J.~M., \&
  {Salvati}, M. 1986, \apj, 301, 790, \dodoi{10.1086/163944}

\bibitem[{{Weiler} {et~al.}(2004){Weiler}, {van Dyk}, {Sramek}, \&
  {Panagia}}]{WEILER04}
{Weiler}, K.~W., {van Dyk}, S.~D., {Sramek}, R.~A., \& {Panagia}, N. 2004, New
  Astronomy Review, 48, 1377, \dodoi{10.1016/j.newar.2004.09.017}

\end{thebibliography}

\listofchanges
\end{document}